\documentclass[a4paper,fleqn]{cas-sc}

\tnotemark[1]
\tnotetext[1]{This is a preprint.}
\renewcommand{\printFirstPageNotes}{} 

\usepackage[numbers]{natbib}

\usepackage{cleveref}

\usepackage{geometry}
\usepackage[
	textsize=scriptsize,
	colorinlistoftodos,
	disable
]{todonotes}

\reversemarginpar
\usepackage{tabularray}

\usepackage{tikz}
\usetikzlibrary{shapes.geometric, arrows, arrows.meta, positioning}

\tikzstyle{block} = [rectangle, draw, fill=blue!20, text centered, rounded corners, minimum height=3em, minimum width=6em]
\tikzstyle{line} = [draw, -latex']

\tikzstyle{startstop} = [rectangle, rounded corners, minimum width=3cm, minimum height=1cm,text centered, draw=black, fill=red!30]
\tikzstyle{process} = [rectangle, minimum width=3cm, minimum height=1cm, text centered, draw=black, fill=orange!30]
\tikzstyle{decision} = [diamond, minimum width=3cm, minimum height=1cm, text centered, draw=black, fill=green!30]
\tikzstyle{arrow} = [thick,->,>=stealth]

\def\tsc#1{\csdef{#1}{\textsc{\lowercase{#1}}\xspace}}
\tsc{WGM}
\tsc{QE}

\begin{document}
\let\WriteBookmarks\relax
\def\floatpagepagefraction{1}
\def\textpagefraction{.001}

\shorttitle{Optimizing AI-Assisted Code Generation: Enhancing Security, Efficiency, and Accessibility in Software Development}    

\shortauthors{Torka et~al.}  

\title [mode = title]{Optimizing AI-Assisted Code Generation: Enhancing Security, Efficiency, and Accessibility in Software Development}  

\tnotemark[1] 

\tnotetext[1]{BMBF, BSI funded} 

%

\author[1]{Simon Torka}

\cormark[1]


\ead{simon.torka@dai-labor.de}

\ead[url]{https://dai-labor.de/}


\affiliation[1]{organization={DAI-Labor, Technische Universität Berlin},
            addressline={Ernst-Reuter-Platz 7}, 
            city={Berlin},
            postcode={10587}, 
            country={Germany}}

\author[1]{Sahin Albayrak}


\ead{sahin.albayrak@dai-labor.de}

\ead[url]{https://www.dai-labor.de}


\cortext[1]{Corresponding author}



\begin{abstract}
\ignorespaces
\label{sec:Abstract}
\todo{https://www.sciencedirect.com/journal/journal-of-systems-and-software/about/call-for-papers => Reliable and Secure Large Language Models for Software Engineering}
In recent years, the rise of AI-assisted code-generation tools has significantly transformed software development. While code generators have mainly been used to support conventional software development, their use will be extended to powerful and secure AI systems. Systems capable of generating code, such as ChatGPT, OpenAI Codex, GitHub Copilot, and AlphaCode, take advantage of advances in machine learning (ML) and natural language processing (NLP) enabled by large language models (LLMs). However, it must be borne in mind that these models work probabilistically, which means that although they can generate complex code from natural language input, there is no guarantee for the functionality and security of the generated code.

However, to fully exploit the considerable potential of this technology, the security, reliability, functionality, and quality of the generated code must be guaranteed. This paper examines the implementation of these goals to date and explores strategies to optimize them. In addition, we explore how these systems can be optimized to create safe, high-performance, and executable artificial intelligence (AI) models, and consider how to improve their accessibility to make AI development more inclusive and equitable.
\end{abstract}




\maketitle


\section{Introduction}
\label{sec:Hintergrund}

Software development is currently undergoing a profound change due to the use of AI-supported code generation tools
\footnote{The potential of this technology is not limited to the field of software development and currently also has enormous potential in areas such as medicine, business, science, finance, law, education, media, and entertainment \cite{Borger.2023, Ray.2023, Sundberg.2023, Wong.2023}.}. While earlier systems aimed to reduce the complexity of conventional software development, new systems based on Large Language Models (LLM) have significantly expanded these capabilities. Advances in context understanding, language generation, and multilingualism have contributed significantly to this development through better human-AI interaction \cite{Borger.2023, ZamfirescuPereira.2023, Ray.2023}.

In the field of software development, systems such as ChatGPT, OpenAI Codex, GitHub Copilot, and AlphaCode now enable the translation of natural language into executable code, making software development much easier. These tools can autonomously create programs, classes, functions, and test cases in different programming languages and far outperform conventional autocompletion tools \cite{Hossain.2023, Perry.2023, Sarkar.2022, Taeb.2024, Mozannar.2024, Wong.2023}. For example, GitHub Copilot generates code that is comparable to human-written code in terms of complexity and readability \cite{AlMadi.2022, Sarkar.2022}. This leads to considerable time savings and increases efficiency, effectiveness, and productivity, which also shortens the time to market for new software \cite{Klemmer.2024, NegriRibalta.2024, Sarkar.2022, Taeb.2024, Improta.2023}. As a result, 70\% of professional developers\footnote{siehe dazu auch Stack Overflow (SO) Developer Survey, 2023 \cite{StackOverflow.20240805}}  use or plan to use AI technologies throughout the entire software development cycle\footnote{The Atlassian ‘State of Developer Experience Report 2024’ shows, however, that developers are less enthusiastic about AI than their managers: 62\% believe that AI assistants contribute little to efficiency, although 61\% see future potential \cite{Atlassian.20240802}.} \cite{Klemmer.2024}. AI-based tools are already being used in areas such as project planning, requirements definition, requirements analysis, code generation, testing, refactoring, change management, optimization, debugging, and documentation as well as in the explanation of code \cite{Hossain.2023, Klemmer.2024, Ray.2023, Res.2024, Sarkar.2022}. However, it can also be increasingly observed that these systems are being used for security-related tasks such as the detection and elimination of vulnerabilities, coding errors, and security gaps, but also for the development of malware and exploits \cite{Berabi.2024, Klemmer.2024}.

The rapid development of these technologies and their enormous potential not only increases the productivity of professional developers but also increasingly opens up the possibility for non-AI/code specialists \cite{Feltus.2021}, private individuals \cite{Sundberg.2023} and non-profit organisations \cite{Kshirsagar.2021} to create programmes and AI models. One outstanding example is ChatGPT, which uses prompting techniques to enable intuitive interaction in natural language, thus transforming conventional programming and AI development in the long term \cite{Ray.2023}.

The social potential can be utilized particularly in the area of ‘AI for Good’ (AI4G)\footnote{AI4G addresses social, environmental, and economic challenges by supporting projects that are not commercially viable but are central to sustainability, health, humanitarian aid, and social justice. It is closely linked to the United Nations Sustainable Development Goals (SDGs) and emphasizes the transformative role of AI in areas that have so far been little penetrated by this technology \cite{Kshirsagar.2021}.} to promote socially effective projects independently and efficiently \cite{Kshirsagar.2021}. Supporting technologies such as LLMs (Large Language Models) facilitate access to AI and software development and promote the use of software and AI in these areas.

Despite these advantages, there are also challenges, particularly concerning the security, reliability, and quality of the generated code \cite{Sarkar.2022}. For example, many users lack the necessary expertise to sufficiently validate the results generated by LLM. In addition, the misuse of these tools, whether intentional or unintentional, can have serious consequences, especially as they are increasingly being used to develop code in safety-critical environments \cite{Klemmer.2024}. 

To overcome these barriers, user-friendly \cite{Yang.2018}  No-/low-code assistance systems \cite{Sundberg.2023}, AI platforms\footnote{e.g. AI4EU \cite{AI4EU.20231114} or its successor ‘AI-on-Demand Platform’ (AIoD) \cite{AIoD.20240806}.} and the introduction of AI ecosystems \cite{Pinhanez.2019} offer promising approaches. These tools enable non-experts to actively participate in knowledge analysis and AI development and thus promote participation in software development processes \cite{Perry.2023}.

This paper aims to analyze current developments in AI-supported code generation, in particular concerning safety aspects, reliability, and code quality. Based on this, a tool is to be developed that generates secure, efficient, and functional code as well as AI models that are also easy to use for non-experts.

Section \ref{sec:Aktueller_Entwicklungsstand} provides an overview of current developments in AI-based code generation tools, in particular the role of LLM and its increasing influence on software development. Section \ref{sec:KI-Codegeneratoren} focuses on the use of AI systems to generate software code and AI models, while section \ref{sec:Herausforderung_und_Risiken_beim_Einsatz_von_KI-Codegeneratoren} covers the technical and security risks associated with these tools. Together, these sections explore how AI code generation can be safely integrated into workflows to maximize benefits and minimize risks. Section \ref{sec:Attack_and_Proteckt_AI_Code_Gen} explores attack vectors such as prompt injection and jailbreaks, as well as mitigation strategies. Section \ref{sec:Attack_and_Proteckt_with_AI_Code_Gen} highlights the dual role of AI code generators as tools for attacking and defending systems. Section \ref{sec:Verbesserungen_für_einen_KI-Codegenerator_Ein umfassender Ansatz} picks up on these insights and presents a system that generates secure and executable code and optimizes human-AI interactions even for non-experts. Finally, section \ref{sec:Fazit} summarises the most important findings of the paper.


\section{Current state of development}
\label{sec:Aktueller_Entwicklungsstand}

AI-based code generators – often referred to as AI assistants – support developers in software development by using machine learning (ML) and natural language processing (NLP) \cite{Hossain.2023}. They are usually based on large language models (LLMs) that enable natural language interaction and can understand and process complex, multi-step instructions \cite{Cotroneo.2024, Hossain.2023, Sarkar.2022, Taeb.2024, ZamfirescuPereira.2023}.

\begin{figure}[htbp]
\centering
\includegraphics[width=0.5\textwidth]{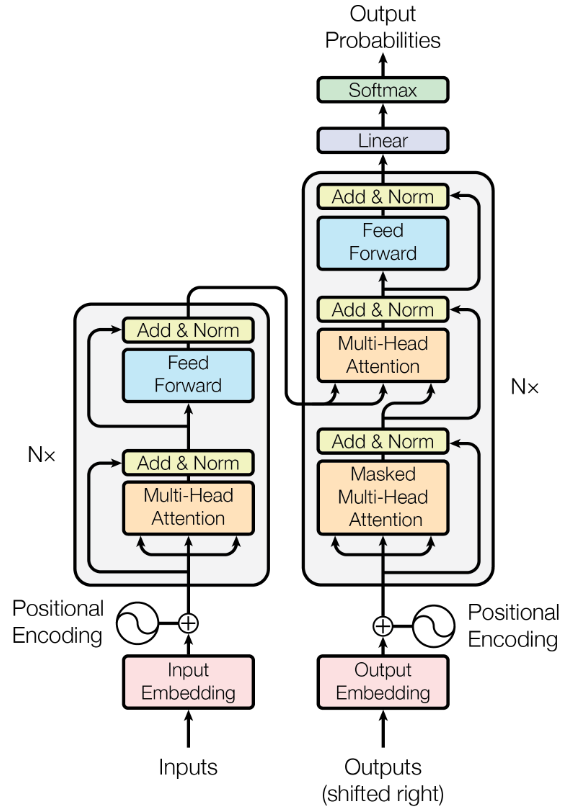}
\caption{The Transformer - model architecture. \cite{Vaswani.2017}.}
\label{fig:LLM_Modelstypes_V2}
\end{figure}

A key technical aspect that significantly shapes the capabilities of these models is their distinction into \textbf{encoders, decoders, and encoder-decoder (sequence-to-sequence)}-models, which perform different tasks in the processing of speech and code \cite[P. 83-86]{tunstall_natural_2023}:

\begin{itemize}
\item \textbf{Encoder models} (like BERT) are specialized in analyzing input sequences and generating context-dependent representations. They are particularly well-suited for analyzing code and detecting patterns or errors (see the left side of Figure \ref{fig:LLM_Modelstypes_V2}).

\item \textbf{Decoder models} (such as GPT-3) focus on generating sequences based on an input, such as a short text or code prompt. They are particularly efficient at creating new code or text by predicting further tokens based on the previous context (see the right side of Figure \ref{fig:LLM_Modelstypes_V2}).

\item \textbf{Encoder-decoder models} (also known as sequence-to-sequence models, such as T5 or BART) combine both approaches by processing input sequences and generating matching output sequences from them. They are particularly useful for transformation tasks, such as translating natural language into code or automating code repairs \cite{Sarkar.2022, Hossain.2023} (see the overall structure in Figure \ref{fig:LLM_Modelstypes_V2}).
\end{itemize}

Another interesting development is the increasing use of platforms such as Hugging Face \cite{hugging_face_hugging_2024} and other LLM hubs, which are central points of contact for pre-trained models and tools for fine-tuning and integrating LLMs \cite{Sundberg.2023}. These platforms provide access not only to models but also to the infrastructure to adapt and use them efficiently for different tasks. This enables a broad application of LLMs in software development and beyond, for example for generating models that meet the requirements of specific domains.

However, the increasing use of such LLM-based assistants also presents challenges. For example, their increasing use endangers previously established information sources such as search engines or online communities such as Stack Overflow, which could displace them \cite{Klemmer.2024, Sarkar.2022, Burtch.2024}. However, since these information sources often also serve as the training basis for the LLM models, a decline in the availability and quality of this data could have negative feedback effects on the performance of the LLM models \cite{Burtch.2024, Klemmer.2024}. Nevertheless, the prevalence of such systems is increasing, as evidenced by telemetry data from the Bing Developer Assistant \cite{Wei.2015, Zhang.2016}. This data shows that programmers use AI-supported tools even if they already know the solution, as these tools make work faster and more efficient \cite{Wei.2015, Zhang.2016}.

\section{AI code generation}
\label{sec:KI-Codegeneratoren}

In artificial intelligence (AI), code generators have established themselves as powerful tools that can accelerate and automate the development of both conventional software \cite{Mozannar.2024, Rajamani.2022, Wong.2023} and the creation of AI systems \cite{Sarkar.2022, Sundberg.2023, Yang.2018}. To provide an overview of the AI systems cited in this work, \cref{tab:Tab_1_Most_commonly_used_Tools_for_AI-Codegeneration}
presents the best-known and most frequently discussed AI code generators while \cref{tab:Tab_2_AI_Tools_used_in_the_analyzed_literature} highlights other AI-based code generators mentioned in the analyzed sources. The following explains how these systems can program conventional code and create AI models.

\begin{table}[htbp]
\centering
\caption{Most Commonly Used Tools for AI Code Generation}
\label{tab:Tab_1_Most_commonly_used_Tools_for_AI-Codegeneration}
\begin{tblr}{
  width = \linewidth,
  colspec = {Q[90]Q[500]Q[165]},
  cells = {l},
  vline{2-3} = {-}{},
  hline{2} = {-}{},
}
Name & Description & Using by \\

ChatGPT & ChatGPT \cite{ChatGPT.20240919}, a product of OpenAI, is one of the most versatile GPT models for code generation. It has been the subject of numerous studies and is constantly being developed further.  & \cite{Cotroneo.2024, Hossain.2023, Improta.2023, NegriRibalta.2024, Taeb.2024} \\

Copilot  & GitHub Copilot \cite{GitHub.20240919}, also developed by OpenAI and provided in collaboration with GitHub, offers developers suggestions directly in the development environment. &  \cite{Cotroneo.2024, Hossain.2023, Improta.2023, NegriRibalta.2024} \\

OpenAI Codex & Codex \cite{Chen.07.07.2021}, a version of GPT-3 specifically designed for code generation, is the basis for GitHub Copilot. It has proven itself in various benchmarks and applications.It can solve an average of 47/164 problems in one attempt in the HumanEval code generation benchmark \cite{Sarkar.2022}. In a dataset with 10,000 programming problems, Codex solves about 3 \% of the problems within 5 attempts \cite{Sarkar.2022}. Major weaknesses are that Codex can generate syntactically incorrect or undefined code and call functions, variables and attributes that are undefined or out of scope \cite{Sarkar.2022}. &  \cite{Hossain.2023,Perry.2023, NegriRibalta.2024, Sarkar.2022, Taeb.2024} \\

AlphaCode & AlphaCode \cite{Li.2022}, developed by DeepMind, was trained directly using GitHub data and optimized for programming competitions. It can reduce a large number of potential solutions to a handful of valid candidates and achieves a success rate of 34 \% in competitions \cite{Sarkar.2022}. This puts it on a par with the average human participant \cite{Sarkar.2022}. In a data set with 10,000 programming problems, AlphaCode solves about 4 - 7 \% of the problems within 5 attempts \cite{Sarkar.2022}.  AlphaCode 2 even performs up to 85 \% better than human competitors in coding tasks \cite{leblond_alphacode_2023, Res.2024}. & \cite{Sarkar.2022}  

\end{tblr}
\end{table}
\begin{table}[htbp]
\centering
\caption{AI Tools used in the analyzed literature}
\label{tab:Tab_2_AI_Tools_used_in_the_analyzed_literature}
\begin{tabular}{>{\centering\hspace{0pt}}m{0.115\linewidth}|>{\centering\hspace{0pt}}m{0.231\linewidth}||>{\centering\hspace{0pt}}m{0.096\linewidth}|>{\centering\hspace{0pt}}m{0.198\linewidth}||>{\centering\hspace{0pt}}m{0.1\linewidth}|>{\centering\arraybackslash\hspace{0pt}}m{0.179\linewidth}}
Name & Used by & Name & Used by  & Name & Used by \\ 
\hline

Amazon CodeWhisperer & \cite{Improta.2023} & Facebooks InCoder & \cite{Chen.07.07.2021, Fried.12.04.2022, Perry.2023} & Mistral Mixtral & \cite{NegriRibalta.2024} \\

CodeBERT & \cite{Li.09.05.2023, NegriRibalta.2024,  Xu.2022} & Google Gemini & \cite{Cotroneo.2024, Gemini.20240806} & PolyCoder & \cite{Li.09.05.2023, NegriRibalta.2024, Xu.2022} \\

CodeGen & \cite{Li.09.05.2023, NegriRibalta.2024, Nijkamp.25.03.2022} & Google PaLM & \cite{Li.09.05.2023, NegriRibalta.2024} & Salesforce CodeGen & \cite{Improta.2023, Li.09.05.2023, NegriRibalta.2024} \\

CodeParrot & \cite{Li.09.05.2023, NegriRibalta.2024} & Gpt2csrc & \cite{NegriRibalta.2024, Pearce.2022} & StarCoder  & \cite{Li.09.05.2023, NegriRibalta.2024} \\

CodeT5 & \cite{NegriRibalta.2024} & IntelliJ IDEA & \cite{Hossain.2023} & Tabnine & \cite{Hossain.2023} \\

DeepAPI-onlySec & \cite{NegriRibalta.2024} & Meta LLaMA & \cite{Li.09.05.2023, NegriRibalta.2024}	& GitHub Copilot & \cite{Improta.2023} \\

DeepAPI-plusSec & \cite{NegriRibalta.2024} & Microsoft Copilot & \cite{Cotroneo.2024, Li.09.05.2023, MicrosoftCopilot.20240806}	&  OpenAI ChatGPT & \cite{Improta.2023} \\

%
%
%
%
%
%

\end{tabular}
\end{table}

\subsection{AI code generation of conventional code}
\label{ssec:KI-Codegeneratoren_für_herkömmliche_Softwareentwicklung}

\begin{figure}[htbp]
    \centering
    \includegraphics[width=1\textwidth]{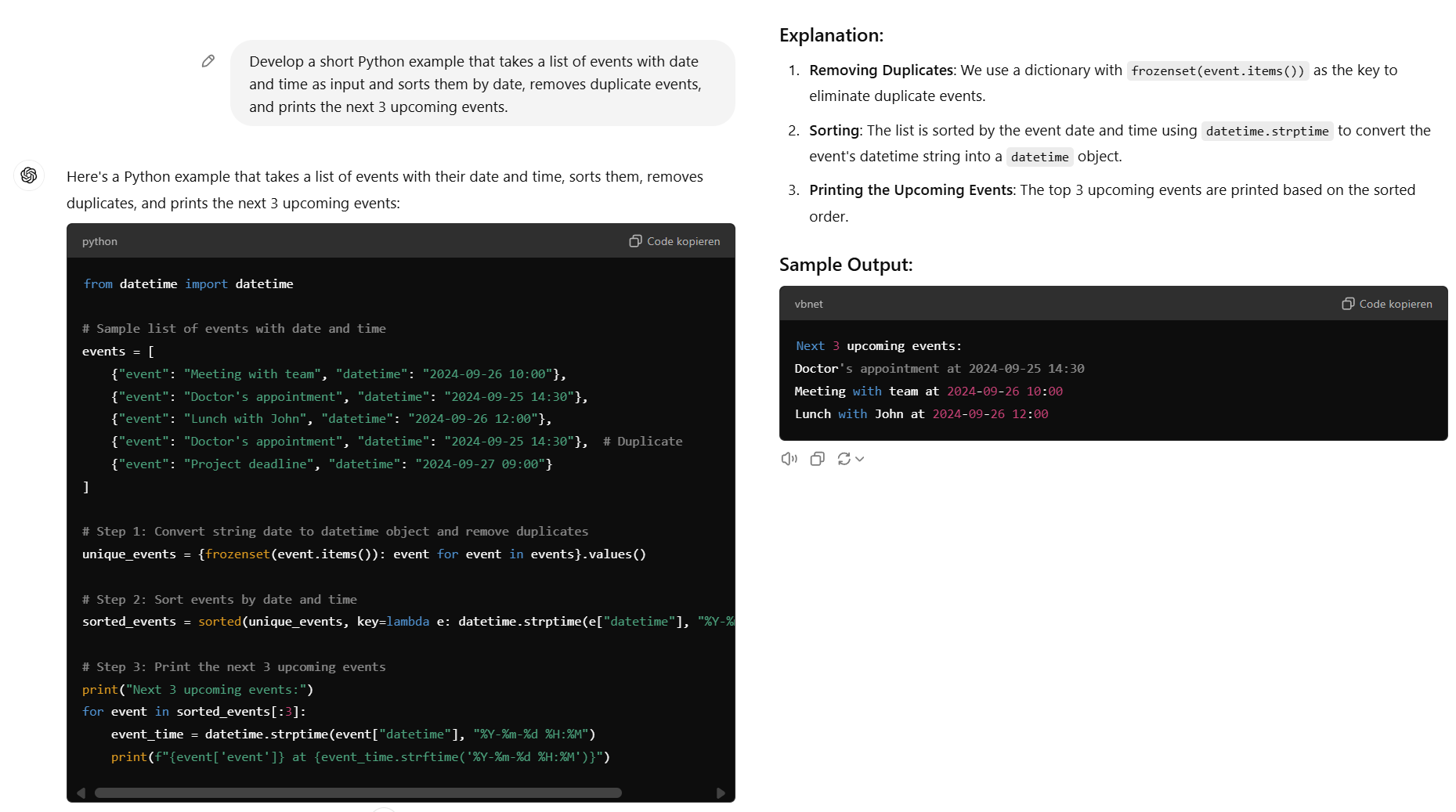}
    \caption{This example shows how ChatGPT effectively solves a simple problem by generating a comprehensive Python script to manage a list of events. The script includes detailed comments to aid understanding and make it easier for non-experts to follow along. ChatGPT also provides a detailed explanation of how the code works and a sample output.}
    \label{fig:ChatGptPromptResponse}
\end{figure}

In software engineering, AI code generators can be used in various ways, e.g. in software development, data processing and analysis, and as cooperative non-human assistants \cite{Ray.2023}. AI-supported functions for code generation, completion, translation, optimization, summarisation, verification, error correction, documentation, and clone detection far outperform existing systems \cite{Mozannar.2024, Ray.2023, Wong.2023}. For example, they can generate program code in various programming languages based on prompts \cite{Mozannar.2024, Wong.2023}. This behavior can be seen in Figure \ref{fig:ChatGptPromptResponse}.

In addition, the use of AI techniques allows knowledge and experience to be derived from previous projects \cite{Wong.2023, Yang.2018}. The high performance of these systems results in particular from the use of modern LLMs that have been trained with billions of data points \cite{Mozannar.2024}. Due to this performance, LLMs have proven to be an important design method that is potentially accessible to non-experts and makes the use of AI systems more intuitive \cite{ZamfirescuPereira.2023}.

On the economic side, the use of such AI techniques improves the efficiency and quality of software-intensive projects \cite{Wong.2023} and accelerates business processes, leading to higher productivity and profitability for companies \cite{Sarker.2022}. On the scientific and private user side, this technology could also give non-experts access to software-supported problem solving \cite{Borger.2023} and thus save time and resources by eliminating the need for external software specialists \cite{Sundberg.2023}.

\citeauthor{Sundberg.2023}~\cite{Sundberg.2023} point out, however, that the responsible use of AI can be a competitive disadvantage for smaller companies, which often do not have the necessary technological skills. In addition, the code generated by AI currently still needs to be carefully validated, as it may be erroneous and contain security vulnerabilities \cite{Sarkar.2022}, which means additional effort. This disadvantage is worse for non-AI experts, as these systems can only be used profitably if expertise in the field of prompt engineering is available \cite{ZamfirescuPereira.2023} \footnote{see also section \ref{sssec:Optimierung_der_Prompts})}. 

To facilitate the accessibility of AI programming assistants for research, business, and private individuals, these systems must be secure, easily accessible, and intuitive to use. \citeauthor{Rajamani.2022}~\cite{Rajamani.2022} presents the vision of a futuristic programming environment based on ML models and logical rules that continuously evolve during use, while \citeauthor{Sundberg.2023}~\cite{Sundberg.2023} propose the development of AI platforms as a solution. By using such platforms, a bridge can be built between different groups of experts (e.g. technical experts and business experts), overcoming existing technical, linguistic, and interest-related barriers and enabling solutions to complex problems to be reached more quickly and efficiently \cite{Sundberg.2023}. In addition, \citeauthor{Sundberg.2023}~\cite{Sundberg.2023} show that the democratization of AI and the associated expansion of the user base can lead to new, diversified, and more innovative AI solutions. 

In addition, the use of AI also gives rise to technical \cite{Ray.2023, Ylitalo.2024}, ethical \cite{Borger.2023, Ray.2023}, legal \cite{Borger.2023, Ray.2023} and social \cite{Borger.2023} requirements that must be analyzed and considered in detail. For example, while these models often produce promising results, they also have considerable risks and limitations. A common problem is that models replicate entire solutions or significant parts of them directly from the training data instead of generating new code \cite{Sarkar.2022}. This raises questions about the quality and originality of the generated code. Furthermore, even developers' efforts to curate and maintain high-quality datasets do not always guarantee that the output is error-free or optimal \cite{Sarkar.2022}. In addition, performance consistency remains a constant challenge \cite{Sarkar.2022}. Furthermore, the installation, maintenance, and management of AI systems represent a further obstacle \cite{Sundberg.2023}.

\subsection{AI code generation of AI models}
\label{ssec:KI-Codegeneratoren_zur_Generierung_von_KI}

\begin{figure}[htbp]
    \centering
    \includegraphics[width=1\textwidth]{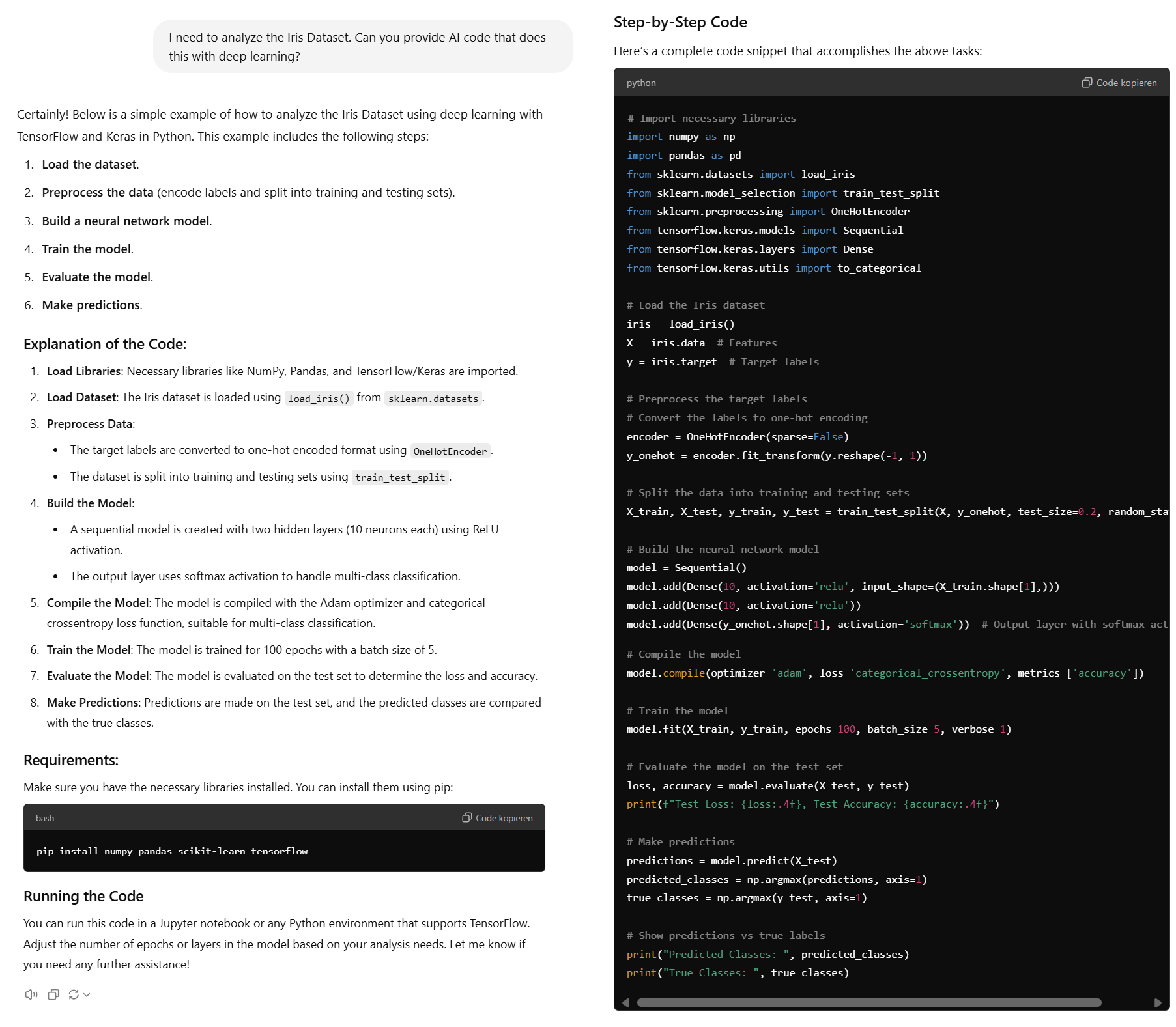}
    \caption{This example shows how ChatGPT effectively solves a problem using deep learning techniques by generating a comprehensive Python script to classify the Iris dataset. The output includes details of the process and code functionality, required libraries, useful side information and the code itself.}
    \label{fig:ChatGptPromptResponseForAI}
\end{figure}

An increasingly important application of AI code generators is the automatic creation of ML and AI code by AI systems, making AI accessible to non-AI experts (An example of how the interaction between ChatGPT and a non-AI expert could look like can be found in Figure \ref{fig:ChatGptPromptResponseForAI}\footnote{Note that ChatGPT already knows the Iris dataset. Dealing with unknown datasets may prove difficult for ChatGPT. In addition, the models developed by ChatGPT based on complex datasets sometimes do not produce good results.}). However, to make AI usable for these non-experts or experts from other domains and to meet the increasing demand for knowledge-intensive processes, the technology for creating such AI models must be easily accessible and secure. \citeauthor{Feltus.2021}~\cite{Feltus.2021} therefore call for AI tools to enter into a symbiotic relationship with the creators and to pursue a user-centric, simple AI generation approach. In particular, this should also promote a balanced interaction between AI-supported innovation and human expertise \cite{Ray.2023}. The approaches already explained for the synthesis of conventional software apply to systems for the automatic synthesis of AI code, extended by the principles following in this section.

However, the development of AI models is often subject to high entry barriers, which makes accessibility more difficult. Therefore, experts are currently required for development \cite{Sarker.2022}. Another hurdle is the validation of models, which can hardly be guaranteed by experts from other domains and laypersons \cite{Yang.2018}. As they often only consider the accuracy of the system, problematic or incorrect models could be used \cite{Yang.2018}. In addition, access restrictions for the development of AI models must be reduced with the help of suitable concepts such as interactive assistance systems to provide the best possible support for these groups \cite{Sarker.2022, Pinhanez.2019, Wong.2023, Yang.2018}.

The introduction of these systems can ensure the democratization of AI technology while guaranteeing easy and user-friendly access \cite{Sundberg.2023}. This can make AI accessible and affordable not only for large companies but also for small businesses and organizations, academia, and individuals. To enable inexperienced users without coding, security, and AI experience to develop, deploy, operate, and train AI models, these no-/low-code AI platforms must be made accessible to this user group through a combination of AI assistants, automated MLOps, and user-friendly graphical interfaces \cite{Sundberg.2023}. This also means that the need for often unavailable experts such as data scientists and ML/AI specialists can be reduced \cite{Sundberg.2023}.

\citeauthor{Ylitalo.2024}~\cite{Ylitalo.2024} point out that such AI tools must also be adapted to the target group. They show this using the example of the study “Github Copilot AI Pair programmer: asset or liability” \cite{MoradiDakhel.2023} which suggests that the GitHub Copilot is a good tool for senior developers, but offers only limited added value for junior developers. In addition, the models created by non-AI experts and laypeople must be prepared in such a way that they can understand and assess the impact of using their code \cite{Feltus.2021}.

To create user-friendly graphical interfaces for no-/low-code programming assistance systems, tools such as “test-driven machine teaching” \cite{Yang.2018} can be used, which supports designers in developing GUIs that adapt to the interaction flow (interactive machine learning (iML)) and the circumstances of laypersons and at the same time point out technical pitfalls that can occur when creating ML models. Through these interactive and user-friendly graphical interfaces, as well as processes adapted to the needs and skills of users in the field of ML/AI and programming, more robust AI models can be developed by non-experts \cite{Yang.2018}. The guided processes that can be integrated into the GUI support compliance with basic rules and safety aspects \cite{Yang.2018} and thus also ensure a safer and more ethical use of AI.

A first approach to solving this problem is provided by \citeauthor{Pinhanez.2019}~\cite{Pinhanez.2019} with the “Massive Open Learning AI System” (MOLA), a community-based and collaborative approach to creating AI development by domain experts without AI knowledge.  The advantage of an open, integrative AI ecosystem results from the leveraging of synergy effects within the community and the availability of expert knowledge from different disciplines. \citeauthor{Pinhanez.2019}~\cite{Pinhanez.2019} investigate the collaborative development of an AI assistant for diabetes patients in their study. This also shows the advantages of such a system. For example, the community noticed that the AI assistant was not well received by some ethnic groups due to cultural differences, which could subsequently be resolved cooperatively. Another AI platform was presented as part of the AI4EU research project \cite{AI4EU.20231114}. AI4EU provides a container-based architecture for the collaborative development of AI models by experts and non-experts. In addition, \citeauthor{Sundberg.2023}~\cite{Sundberg.2023} highlight cloud-based AI platforms, so-called “AI as a service” systems, as a potential new economic sector. Such a cloud-driven approach makes it possible to build an extensive AI ecosystem that includes both lay and expert users \cite{Sundberg.2023}. It should also be noted that AI models are considered trade secrets in many companies and therefore must not be made publicly accessible.


However, a major problem in the development of such systems is the need for extensive and high-quality training data. One source of data that has been largely ignored to date is the online platform Kaggle, which was examined by \citeauthor{Bojer.2021}~\cite{Bojer.2021}. In their study, they analyzed six different Kaggle competitions and contrasted the claims and quality of the solutions with the M-competitions\cite{UniversityofNicosia.20240329} organized by Spyros Makridakis and Michèle Hibon. Their investigation focused on forecasting problems, such as weather data and production and sales forecasts. To validate the performance, they created benchmarks for these six competitions based on simple and proven naive and seasonal naive methods, which are then compared with the Kaggle solutions. The best solutions on Kaggle outperformed the simple solutions by a minimum of 25 \% and a maximum of 74 \%. Even the solutions in 50th place outperformed the simple solutions by at least 10 \%. In addition, \citeauthor{Bojer.2021}~\cite{Bojer.2021} show that the criticism leveled at the M-competitions, such as the lack of high-frequency and hierarchical time series as well as the lack of secondary information such as exogenous variables, which makes these competitions less representative of real economic problems \cite{Darin.2020, Fry.2020}, is partially resolved in the Kaggle competitions. Kaggle competitions thus offer a good learning opportunity and prove the hypothesis put forward in the M competitions that combinations of different ML tools perform better than individual solutions. 

One shortcoming of Kaggle, however, is that the platform only requires solutions, and participants are not obliged to upload the associated code, although the best solutions usually also have source code. This underlines the importance of the solutions submitted on Kaggle and their potential as a training corpus for LLM-based AI code generators.

Concerning the development of knowledge sources, however, \citeauthor{Sarkar.2023}~\cite{Sarkar.2023} points out that the integration of structured (e.g. knowledge graphs) and unstructured knowledge (e.g. Wikipedia) represents a considerable challenge. In addition, it is necessary that the knowledge model can be understood and used by both a machine (AI) and a human \cite{Feltus.2021}.

Due to these developments, the use of LLMs has proven to be an important design method that is also accessible to non-AI experts and makes the use of AI systems more intuitive. Through a symbiotic relationship between technology and creators, robust and ethical AI models can be developed that can be understood and used by both machines and humans. Research into user-friendly and flexible no-/low-code AI generators promotes the acceptance of AI and at the same time enables the integration of AI in all areas of society \cite{Feltus.2021}.

\section{Challenges and risks when using AI-based code generation}
\label{sec:Herausforderung_und_Risiken_beim_Einsatz_von_KI-Codegeneratoren}

With the increasing integration of AI-based code generators into the software development process, it is becoming increasingly important to understand the associated challenges and risks. While high-performance models can have a positive economic impact for the organizations using them \cite{Sarkar.2022, Klemmer.2024, Improta.2023}, inferior models can both reduce productivity and pose significant security risks \cite{Improta.2023}. In particular, the introduction of programming wizards adapted to end users poses specific challenges that affect professional programmers and amateurs alike \cite{Sarkar.2022}. 

\subsection{Technical challenges}
\label{ssec:Technisch_bedingte_Herausforderungen}

On the technical side, there are particular requirements in the areas of reliability, accuracy, bias control, generalization capability, explainability of the models, contextual understanding and safety aspects \cite{Ray.2023}, which in turn can have an impact on the areas of ethics, law and social issues, as the example of incorrect or ethically questionable answers shows. This is in particular due to the training data and the technical properties of LLMs. 

Since collecting training data is a time-consuming and expensive process \cite{Improta.2023}, the datasets used for training are typically collected by crawling billions of lines of source code \cite{Cotroneo.2024, Mozannar.2024} from publicly available, voluminous, unverified, sometimes incomplete, and not always trustworthy online sources\footnote{According to \citeauthor{Wong.2023}~\cite{Wong.2023}, this could be “online software artifacts such as source code repositories [e.g. GitHub \cite{Cotroneo.2024} and StackOverflow threads \cite{Improta.2023}], bug databases and code snippets”.}. In addition, it is often unknown how secure or efficient this code is. As a result, the models trained with this data also learn to generate the weaknesses present in the training data, such as not taking into account important edge cases, using outdated and insecure functions and libraries as well as exploiting incomplete, non-executable, insecure, security-critical, faulty or even malware-compromised code \cite{Cotroneo.2024, Improta.2023, Klemmer.2024, NegriRibalta.2024, Perry.2023, Sarkar.2022}. 

In addition, these models can only answer correctly as long as the question is within the scope of their training data, otherwise, the models give unclear or incorrect answers or start to hallucinate \cite{Klemmer.2024, NegriRibalta.2024}. Furthermore, some models include previous questions and answers within a session in the generation of the output. This can cause the quality of the answers to deteriorate over time. In addition to the training data, the LLM technologies themselves also have their weaknesses. For example, the generated answers are based only on statistical abstractions of the training data \cite{Chen.07.07.2021, NegriRibalta.2024, Vaswani.2017}. These systems can therefore only predict which token is most likely to occur next, and not whether this makes semantic sense \cite{NegriRibalta.2024}. To make matters worse, it is very difficult to detect errors in the training data or the design of the model since the internal process of the LLM/GPT is beyond any quality control \cite{Cotroneo.2024}. Thus, despite all efforts of the developers to use only high-quality and cleaned data to train these generators, no guarantee can be given for its correctness and consistency \cite{Sarkar.2022}. However, improving code quality and security has so far proved difficult, as this requires large amounts of high-quality and, at best, labeled training data \cite{Berabi.2024}. the quality of the answers can deteriorate over time, and complex and non-executable solutions can be output \cite{Klemmer.2024}.

The operation of such extensive AI assistants also requires complex technical knowledge. In addition to DevOps and MLOps, cloud infrastructures are often required to operate modern AI systems. However, the administration and operation of such infrastructures is extremely complex. \citeauthor{Ylitalo.2024}~\cite{Ylitalo.2024} offer a solution by investigating how generative artificial intelligence can support the management and operation of cloud infrastructure. They rely on an AI-automated “infrastructure as code” approach, which can improve service quality and simplify the necessary processes. However, the authors focus on supporting cloud experts in the management of such systems and not on replacing them.

For the reasons mentioned above, companies and their developers are still reluctant to make widespread use of this technology \cite{Res.2024}. Therefore, future work should focus on further improving the security and accuracy of these tools to ensure safe use \cite{Taeb.2024}.

\subsection{Code quality}
\label{ssec:Codequalität}

A crucial aspect when evaluating AI code generators, in addition to correct functionality, is the security of the generated code. According to Open Source Security and Risk Analysis (OSSRA), 40\% of AI-generated code has security vulnerabilities and is considered unsafe\cite{Hossain.2023, Pearce.2022, Synopsys.20240611}. The performance of AI code generators depends on factors such as prompt, programming language, application domain, and type of vulnerability \cite{Pearce.2022, Pearce.2023}\footnote{For example, an analysis showed that 50.29\% of the generated C code had security vulnerabilities, while only 38.35\% of the Python code had similar problems \cite{Pearce.2022, NegriRibalta.2024}.}. However, it turns out that the use of AI assistance systems only causes 10\% more security vulnerabilities than a control group without such systems~\cite{Sandoval.2023}. In addition, the choice of code generator itself also influences code security \cite{Taeb.2024, NegriRibalta.2024, Perry.2023, Cotroneo.2024}.

To validate code security, \citeauthor{Taeb.2024}~\cite{Taeb.2024} test the CodeBERT, GPT 3.5, and CodeX systems with code compromised by CVE, CWE, NIST, and NVD vulnerabilities, as well as the top 10 OWASP web vulnerabilities, and analyze how well these systems can repair the compromised code and generate bug-free code. The output of the models was then analyzed by static code analyzers such as Clang-Tidy, FlawFinder and VCG. CodeX was found to have the highest code generation capability, producing accurate, secure and privacy-compliant code. GPT 3.5 showed lower code generation capabilities compared to CodeX, but was very effective in explaining potential vulnerabilities and commenting on code and log files. CodeBert was also able to generate high-quality code but was weaker in terms of security measures. They also show that the quality and security of the output depends on the complexity of the tasks and the accuracy of the instructions and thus, among other things, on the skills of the operator \cite{Taeb.2024}. 

Although not directly examined by \citeauthor{Taeb.2024}~\cite{Taeb.2024}, the study also suggests that the architecture of the model influences the results. CodeBert is based on BERT's encoder architecture, but has been specifically optimized for processing source code and natural language. This architecture allows CodeBert to extract contextual information from the input text and understand semantic meanings. In contrast, GPT-3.5 and Codex are part of the Transformer architecture, which is designed for sequential text generation. While GPT-3.5 uses a broad and diversified text base, Codex was specifically trained on large-scale programming data. This focus improves Codex's ability to generate high-quality code and answer specific programming queries.

In addition to the studies by \citeauthor{Perry.2023}~\cite{Perry.2023}, \citeauthor{Sarkar.2022}~\cite{Sarkar.2022} and \citeauthor{Klemmer.2024}~\cite{Klemmer.2024} already presented in chapter \ref{ssec:User_Studies_Security_and_Trust}, \citeauthor{NegriRibalta.2024}~\cite{NegriRibalta.2024}, \citeauthor{Ylitalo.2024}~\cite{Ylitalo.2024} and \citeauthor{Improta.2023}~\cite{Improta.2023} also analyze the effects of the steadily increasing use of AI assistants on code quality and code security. Based on a systematic literature review, \citeauthor{Ylitalo.2024}~\cite{Ylitalo.2024} and \citeauthor{NegriRibalta.2024}~\cite{NegriRibalta.2024} conclude that current AI code generators have known and easily avoidable vulnerabilities or even malware in their output. \citeauthor{Ylitalo.2024}~\cite{Ylitalo.2024} also emphasize that some of the code blocks proposed by the AI assistance system “Github Copilot” are incomplete or incorrect and also contain serious security vulnerabilities. However, detecting security vulnerabilities in this partially unfinished and non-executable code poses a significant challenge that can only be solved by analyzing complex patterns and distant relationships within the entire generated code \cite{Berabi.2024, Cotroneo.2024}. A study by \cite{Improta.2023} also shows that the code generated by tools such as "GitHub Copilot", "Amazon CodeWhisperer", "Salesforce CodeGen" and "OpenAI ChatGPT" often has security issues and insufficient code quality. Insecure encryption methods and other vulnerabilities in the code are particularly problematic as they facilitate potential attacks. It is therefore considered crucial to develop mechanisms that can detect and fix such security gaps at an early stage \cite{Improta.2023}.

In addition, \citeauthor{Cotroneo.2024}~\cite{Cotroneo.2024} point out that the human factor has a significant impact on the quality and security of code generated by code generators. For example, inexperienced users are usually unable to assess the security of AI-generated code \cite{Cotroneo.2024}. They also note that manual review of AI-generated code cannot be implemented due to the expected increase in its volume \footnote{Currently, manual review is the preferred method of experts such as \cite{Liguori.2021, Liguori.2022}.} \cite{Cotroneo.2024}. They also point out that the use of static analysis tools is usually not possible when the code is incomplete \cite{Cotroneo.2024}. Furthermore, the need for manual verification and the potential presence of serious security risks significantly reduces the adoption of this technology \cite{Cotroneo.2024}.

\subsection{Challenges in prompt engineering}
\label{ssec:Herausforderungen_beim_Prompt_Engeneering}

As previously explained, prompt engineering is one of the biggest challenges in the use of AI code generators \cite{Sarkar.2022, Klemmer.2024}. Writing effective prompts for AI models is difficult and requires detailed instructions and clear logic \cite{Sarkar.2022, Klemmer.2024}. In addition, problems often need to be broken down into sub-problems to be solved by today's AI systems \cite{Sarkar.2022}. This means that users have to specify their intentions in detail and at a fine granular level and often refine them afterward to achieve the desired results \cite{Sarkar.2022}. This is a key obstacle, especially for inexperienced users who have no programming skills but could be significantly relieved by natural language programming assistants \cite{Sarkar.2022}. The interaction interfaces of these systems must therefore be improved in the future in terms of user-friendliness and quality of results \cite{Klemmer.2024}.

\subsection{Assessing the security and trust of AI code generators based on user studies}
\label{ssec:User_Studies_Security_and_Trust}

AI code generators support developers in all areas of the software development cycle \cite{Perry.2023}. It is important that these tools avoid potential security vulnerabilities and generate robust, secure code \cite{Taeb.2024}. To assess the advantages and disadvantages of these systems from the user's perspective, we analyse studies by \citeauthor{Perry.2023}~\cite{Perry.2023}, \citeauthor{Sarkar.2022}~\cite{Sarkar.2022} and \citeauthor{Klemmer.2024}~\cite{Klemmer.2024}.

\citeauthor{Perry.2023}~\cite{Perry.2023} conducted a laboratory study in which two groups worked on security-related tasks in the areas of encryption, access control, databases, data input and output, data types and web technologies, with one group working with the help of AI code generators and the other without their help. The generated code was then evaluated and interviews were conducted with the participants to determine their trust in the technology. To ensure reproducibility, the course of interaction (queries, interaction behavior, model responses) with the AI system and the subsequent safety assessments were recorded anonymously and can be downloaded at \cite{neilaperry_neilaperry-users-write-more-insecure-code--ai-assistants_2024}. Analyzing the data showed that participants with AI assistants produced unsafe code even though they thought it was safe. Inexperienced users tend to consider the generated code to be of high quality without scrutinizing it critically. Experienced developers, on the other hand, are better able to assess security and improve code quality by making targeted adjustments to the input, which ultimately increases their productivity. Participants without AI assistants, on the other hand, wrote safer code and had a more realistic assessment of code quality. 

\citeauthor{Sarkar.2022}~\cite{Sarkar.2022} analyze how LLM-supported programming assistants influence code development based on experience reports from publicly available sources and also found that inexperienced users often tend to blindly trust the results generated by the AI, which poses potential safety risks. In addition, the automation of code generation can lead to less time being invested in testing and understanding.

\citeauthor{Klemmer.2024}~\cite{Klemmer.2024} analyzed 27 interviews and 190 Reddit posts and found that despite safety concerns, the use of AI assistants is increasing in practice. However, the study emphasizes that developers have to subject the code generated by the AI to additional checks, which negates some of the time benefits as this process has been little researched and standardized to date \footnote{Other study participants, however, stated that they had already integrated complex peer review processes and extensive software tests into their AI-supported work process\cite{Klemmer.2024}.}. This could lead to a creeping habituation effect in which the generated code is adopted unchecked, which jeopardizes software security in the long term and makes it more difficult to understand and maintain the code \cite{Klemmer.2024, Sarkar.2022}. To ensure security, AI generators are required that already have mechanisms for generating secure code.

Despite security and quality concerns, AI assistants are used by developers almost every day, although more than half fear negative effects on software security \cite{Klemmer.2024}. This is particularly because AI code generators promise significant productivity increases \cite{Dohmke.2023, Dohmke.26.06.2023, Sarkar.2022, Klemmer.2024, Perry.2023} and are becoming increasingly crucial for competitiveness \cite{Klemmer.2024}. It is estimated that tools such as GitHub Copilot could increase global productivity by 30 \% and boost GDP by USD 1.5 trillion by 2030 \cite{Dohmke.2023, Dohmke.26.06.2023, Klemmer.2024}.

This transformation will increasingly shift developers' responsibilities to the areas of prompt engineering and AI management. However, prompt engineering is a challenge, especially for inexperienced users \cite{Klemmer.2024, Sarkar.2022}.

The key results of the studies are summarised in Table \ref{tab:tab_9_user_study_results}.
\begin{table}[htbp]
    \centering
    \caption{Summary of the results of studies on AI-supported development tools}
    \label{tab:tab_9_user_study_results}
    \begin{tabular}{|l|p{10cm}|}
        \hline
        \textbf{Source} & \textbf{Results and statements of the participants} \\ 
        \hline
        \citeauthor{Klemmer.2024}~\cite{Klemmer.2024} &
        \begin{itemize}
            \item AI assistants are gaining popularity despite concerns about security and quality.
            \item AI code generators are used almost daily, while over half of developers fear negative effects on software security.
            \item AI assistants are considered to be relevant to competition.
            \item The use of AI assistants is crucial for competitiveness.
            \item Increases in productivity reduce personnel costs, making the technology more attractive from a business perspective.
            \item The performance of AI assistants will continue to improve over the next three to four years.
            \item Tasks are increasingly shifting towards prompt engineering and AI management.
            \item Inexperienced users have difficulty formulating prompts.
            \item Participants who used AI assistants produced less secure code.
            \item Participants who used AI assistants believed they wrote more secure code.
            \item Participants blindly trust AI, which can increase security risks.
            \item Generated answers must be extensively checked, which is seen as a disadvantage.
            \item Some participants would like to see further research into simple checking processes for AI-generated content.
            \item Others have already integrated complex peer review processes and extensive software testing into their AI-powered workflows.
        \end{itemize} \\
        \hline
        \citeauthor{Dohmke.2023}~\cite{Dohmke.2023}, \citeauthor{Dohmke.26.06.2023}~\cite{Dohmke.26.06.2023} & 
        \begin{itemize}
            \item AI-powered tools like GitHub Copilot could increase productivity by 30\% and add \$1.5 trillion to global GDP by 2030.
            \item Users accept 30\% of proposed code on average
            \item Productivity increase for less experienced developers.
            \item The innovative power of generative AI is deeply rooted in the open source ecosystem and is revolutionizing software development.
        \end{itemize} \\
        \hline
        \citeauthor{Sarkar.2022}~\cite{Sarkar.2022} & 
        \begin{itemize}
            \item Difficulties in formulating prompts, especially for inexperienced users.
        \end{itemize} \\
        \hline
    \end{tabular}
    \label{tab:ai_studies}
\end{table}
\subsection{Social, legal and ethical challenges}
\label{ssec:Soziale_Rechtliche_und_Ethische_Herausforderunge}

The social, legal, and ethical requirements concern the security and protection of data, the impact on the environment, the transparency and accountability of AI systems as well as the prevention of misuse and the promotion of understandable and legally, ethically and socially acceptable interaction \cite{Ray.2023}.

\citeauthor{Borger.2023}~\cite{Borger.2023} analyze in their work the challenges that arise from the use of LLMs such as ChatGPT and Bard in the areas of scientific research and education concerning these three central areas of tension. It must be noted, for example, that these systems can be used immediately by anyone and can in principle also be misused for malicious purposes that were previously inaccessible to the person using them \cite{Borger.2023}. To protect against misuse, OpenAI, for example, has implemented security measures to prevent malicious use of ChatGPT \cite{Borger.2023}.

There are also social implications that can arise from the use of AI-supported systems. \citeauthor{Borger.2023}~\cite{Borger.2023} show that there are still problems to be solved in the areas of privacy, confidentiality and security. On the other hand, they also show that the use of AI can overcome previous barriers and access restrictions. An example from biology shows the resulting advantages. In the future, biologists will be able to use AI systems for software development, meaning that theoretically any biologist can become a bioinformatician. 

A similar analysis is conducted by \citeauthor{Wong.2023}~\cite{Wong.2023}, who concludes that AI-assisted programming enhances the efficiency and effectiveness of software developers. However, he also emphasizes the need to balance increased productivity with ensuring security, safety, and reliability in the software development process. In addition, he points out that AI-assisted programming can more easily detect critical incidents during the software development lifecycle before they become critical\cite{Wong.2023}. The capability to predict errors can be used in the development of AI code generators to ensure a balance between high productivity and high security by proactively detecting and fixing errors.

\citeauthor{Ylitalo.2024}~\cite{Ylitalo.2024} also point out the ethical aspect. They suggest that the models used and the expenditure of these must be understood in essence to be able to better classify the ethical risks and effects arising from them.

Another point to consider from an ethical perspective can be derived from the findings of \cite{ZamfirescuPereira.2023}, who elaborates that interaction with LLM models requires a certain degree of language and prompting expertise and is not yet oriented towards the needs of non-programmers and end users \cite{ZamfirescuPereira.2023}. It is pointed out that while users can explore these systems opportunistically, they struggle to systematically interact with the system and achieve robust results. \citeauthor{ZamfirescuPereira.2023}~\cite{ZamfirescuPereira.2023} also point out that most problems arise because people also interact with the system as they would with a human, but the cognitive abilities required for this are not yet covered by such systems, which can lead to ethically, legally or socially questionable interactions. \citeauthor{Ray.2023}~\cite{Ray.2023} add to this statement and list various challenges in the development of LLM-based chatbots as problematic. These include maintaining context and context switching, dealing with ambiguity, personalization, the absence of “common sense”, emotional intelligence, ethical considerations, robustness and security, dealing with requests outside of the training data, and scalability and efficiency.

Furthermore, \citeauthor{ZamfirescuPereira.2023}~\cite{ZamfirescuPereira.2023} point out that the development of LLM-based chatbots, such as bots in customer service, is a challenge for non-experts, as their implementation requires both effective strategies for understanding the context and an effective strategy for detecting errors and evaluating system effectiveness. To address this problem, the authors present “BotDesigner” \cite{ZamfirescuPereira.2023}, a no-code-based chatbot designer that enables non-experts to develop a chatbot based on an LLM using prompts alone and to generate a chatbot that guides users through the activity to be performed in an efficient and easy-to-understand manner. 

There are also some legal hurdles. For example, effective protection of privacy and data must be guaranteed following the applicable laws and regulations \cite{Ray.2023}. Another point is the guarantee of copyright. Both the training of models and the actual model output can lead to copyright infringements \cite{Klemmer.2024}. If, for example, the prompts entered and the resulting answers are included in future model training, future models may learn this knowledge and claim to be the source of this information, as metadata from previous sessions, such as the author's name, is usually not available and therefore not included in the training. In this case, it could prove difficult to prove authorship. Furthermore, the answers provided by the assistant may themselves be protected by copyright, which represents an immense legal risk for users \cite{Klemmer.2024}.

\section{Attack possibilities and protection of AI code generators}
\label{sec:Attack_and_Proteckt_AI_Code_Gen}

This section examines how LLMs can be attacked, how malicious actors can abuse LLMs for cyberattacks, and what safeguards can be put in place against such abuse. As the prevalence of LLMs continues to grow, understanding the potential threats and safeguards is essential. A general concept for preventing the attacks described in this chapter is provided in Figure \ref{fig:fig_Attack_Defend_Relations}. In addition, Table \ref{tab:tab_3_Attack_types_And_Counter_Messures} presents some common attacks and their corresponding countermeasures. In addition, the Table shows the process and interactions of attacking and defending such systems.

\begin{figure}[!ht]
\centering
\caption{Relations between Common Attacks and Mitigations on AI Code Generators}
\label{fig:fig_Attack_Defend_Relations}

\begin{tikzpicture}[
    node distance=2cm, 
    attack/.style={rectangle, draw=red!70, fill=red!20, thick, text width=3.5cm, align=center, minimum height=2em},
    defense/.style={rectangle, draw=blue!70, fill=blue!20, thick, text width=3.5cm, align=center, minimum height=2em},
    arrow/.style={-{Latex[length=3mm]}, thick}
]

\node[attack] (reversepsychology) {Reverse Psychology Attack};
\node[attack, below of=reversepsychology, node distance=2cm] (promptinjection) {Prompt Injection Attack};
\node[attack, below of=promptinjection, node distance=3cm] (datapoisoning) {Data Poisoning};
\node[attack, below of=datapoisoning, node distance=3cm] (modelextraction) {Model Extraction};
\node[attack, below of=modelextraction, node distance=2cm] (backdoor) {Backdoor Attack};
\node[attack, below of=backdoor, node distance=1cm] (modelinversion) {Model Inversion Attack};
\node[attack, below of=modelinversion, node distance=2cm] (membershipinference) {Membership Inference Attack};
\node[attack, below of=membershipinference, node distance=2cm, yshift=1cm] (modeltheft) {Model Theft};

\node[defense, right=5.5cm of reversepsychology] (monitoring) {Monitoring and Filtering Input};
\node[defense, right=5.5cm of promptinjection] (promptopt) {Prompt Validation Guidelines};
\node[defense, right=5.5cm of datapoisoning, yshift=0.5cm] (dataverification) {Data Integrity Checks};
\node[defense, below of=dataverification, node distance=1cm] (codeanalysis) {Static and Dynamic Code Analysis};
\node[defense, right=5.5cm of modelextraction] (modsecure) {Model Security Checks};
\node[defense, right=5.5cm of backdoor, yshift=-0.5cm] (accesscontrol) {Access Control Mechanisms};
\node[defense, right=5.5cm of modelinversion, yshift=-2.5cm] (modelaccess) {Model Access Limitation};

\draw[arrow] (reversepsychology) -- (monitoring) node[midway, above, sloped] {Monitors};
\draw[arrow] (promptinjection) -- (promptopt) node[midway, above, sloped] {Mitigates};
\draw[arrow] (datapoisoning) -- (dataverification) node[midway, above, sloped] {Detects and Cleans};
\draw[arrow] (datapoisoning) -- (codeanalysis) node[midway, above, sloped] {Identifies Issues};
\draw[arrow] (modelextraction) -- (modsecure) node[midway, above, sloped] {Prevents};
\draw[arrow] (backdoor) -- (accesscontrol) node[midway, above, sloped] {Mitigates};
\draw[arrow] (membershipinference) -- (modelaccess) node[midway, above, sloped] {Limits};
\draw[arrow] (modelinversion) -- (accesscontrol) node[midway, above, sloped] {Prevents Access};
\draw[arrow] (modeltheft) -- (modelaccess) node[midway, above, sloped] {Limits Queries};
\draw[arrow] (reversepsychology) -- (promptopt) node[midway, above, sloped] {Guidelines};

\node[above=1cm of reversepsychology, align=center] {Attack Vectors on AI Code Generators};
\node[above=1cm of monitoring, align=center] {Defense Mechanisms};

\end{tikzpicture}
\end{figure}
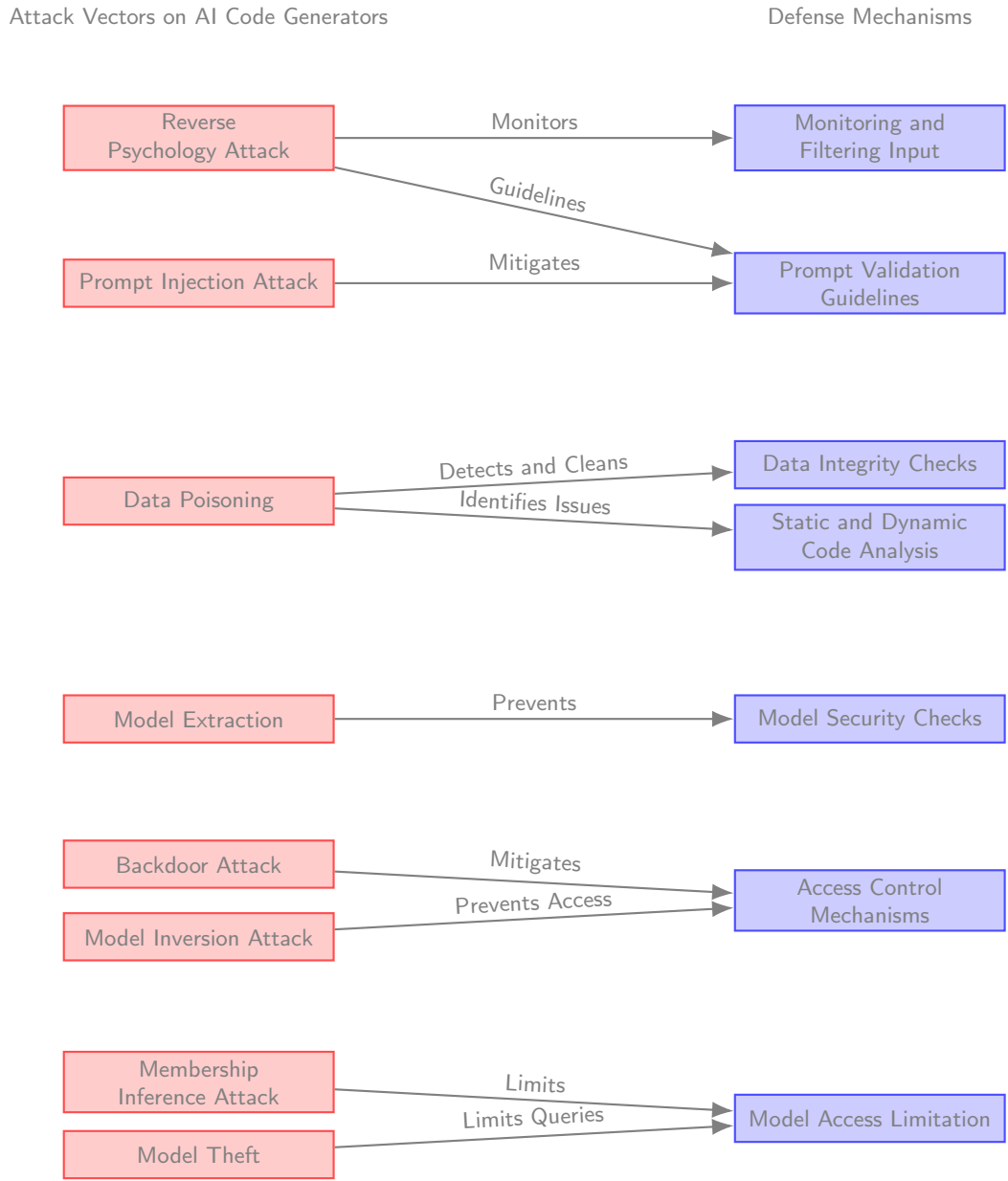
\begin{table}[htbp]
\centering
\caption{Common Security Risks in AI-Assisted Code Generation and Corresponding Mitigation Strategies}
\label{tab:tab_3_Attack_types_And_Counter_Messures}
\begin{tabular}{|>{\raggedright\arraybackslash}p{3cm}|p{5cm}|p{4.5cm}|}
\hline
\textbf{Risk} & \textbf{Description} & \textbf{Solutions} \\
\hline
\textbf{Untrusted Training Data Sources \cite{Cotroneo.2024, Improta.2023}} & Use of unverified or insecure code sources. & Only utilize verified sources and use code analysis tools. \\\hline
\textbf{Prompt Injection \cite{Gupta.2023}} & Malicious inputs trick the AI into generating harmful or unexpected outputs. & Use input validation, filtering, and sanitizing techniques. \\
\hline
\textbf{Data Poisoning \cite{Cotroneo.2024, Improta.2023}} & Attackers inject faulty, manipulated, or biased training data, to alter model behavior. & Use data verification, track sources, and apply anomaly detection. \\
\hline
\textbf{Jailbreak Attacks \cite{Gupta.2023, Ng.20230806}} & Inputs bypass security controls to access restricted features. & Regularly update security measures and implement strict I/O filters. \\
\hline
\textbf{Outdated Libraries \cite{Cotroneo.2024, Perry.2023}} & Vulnerabilities from using old software libraries. & Regularly update libraries and run vulnerability scans. \\
\hline
\textbf{Ethical Decisions \cite{Gupta.2023}} & AI actions can breach ethical standards. & Define ethical guidelines and implement explainable AI for transparency. \\
\hline
\end{tabular}
\end{table}

\subsection{Options for attacking AI generators}
\label{ssec:Angriffsmöglichkeiten_auf_KI-Generatoren}

As AI-assisted code generation becomes more widespread, the number and variety of attack vectors targeting these systems are also growing. Understanding potential vulnerabilities and taking preventive measures is crucial to ensuring the security and reliability of such models. This chapter presents exemplary attack vectors on LLM-based code generators and their impact on the security of the generated code. This analysis is intended to highlight the importance of proactive defense mechanisms.

Among the most significant threats to LLM-based code generators are \textbf{reverse psychology attacks, prompt injection, and jailbreaks}. These techniques make it possible to circumvent the security mechanisms of language models by using specially manipulated inputs to generate unwanted or malicious output \cite{Gupta.2023}. Such manipulations can be subtle and difficult to detect \cite{papernot_limitations_2016}. For example, LLM code generators can be abused to exploit existing vulnerabilities \cite{Gupta.2023, schwinn_adversarial_2023}, to create malware and conduct attacks, or to generate code that contains vulnerabilities \cite{Gupta.2023}. This type of attack emphasizes the need to monitor LLM inputs and outputs carefully.

Furthermore, these attack vectors enable the \textbf{undermining of social and ethical constraints as well as data protection mechanisms} \cite{Gupta.2023}, posing significant risks to both data security and user protection. This also poses the risk of LLM-based code generators unintentionally disclosing sensitive information, as their models usually contain confidential information used during training (so-called model inversion attacks) \cite{fredrikson_model_2015}. Studies show that identifiable (28.2‰), private (7.8‰), and secret (6.4‰) information, such as addresses, medical data, or passwords, can be leaked \cite{NegriRibalta.2024, Niu.2023}. In addition, such leaks pose a significant threat to protecting user data, intellectual property, and corporate information, which can compromise infrastructure and financial assets \cite{Hossain.2023}.

In addition, \textbf{membership inference attacks} pose a further threat. These attacks aim to determine whether certain data points are contained in a model's training dataset, which can have serious consequences, especially when processing sensitive data \cite{shokri_membership_2017}. Such attacks could reveal confidential code snippets or internal company information that greatly benefits organizations. As LLMs are often trained with large amounts of data from corporate sources or open-source code, there is a risk that trade secrets or internal processes could be inadvertently exposed.

Another serious risk to the security and reliability of LLM-based code generators are attacks that specifically target the manipulation of the models, such as \textbf{poisoning and backdoor attacks}. Both types of attacks aim to compromise the functionality of a model by injecting malicious or erroneous data during the training process.

Poisoning attacks aim to deliberately infect the training set of a model with malicious data to affect the quality of the generated code. This can happen both in the pre-training phase and the fine-tuning phase \cite{Improta.2023}. Attackers use two main methods here: In white-box attacks, they gain direct access to the training data to manipulate it in a targeted manner \cite{Improta.2023}. In contrast, black-box attacks, attempt to subvert data collection by web crawlers by injecting erroneous information into the sources used to create the model \cite{Improta.2023}. Backdoor attacks are a particularly insidious extension of this threat. Here, a kind of ‘backdoor’ is integrated into the model during the training phase. This backdoor remains hidden during normal use of the model and is only activated by certain specialized inputs. As soon as it is triggered, the model can generate dangerous or malicious code \cite{gu_badnets_2019}. The challenge in detecting such attacks is that the hidden backdoors are only activated under specific conditions, which makes them extremely difficult to detect. These attacks could compromise LLM-based code generators in such a way that they are later exploited by attackers to generate malicious output \cite{schwinn_adversarial_2023}. Both poisoning and backdoor attacks pose serious threats to the security and integrity of AI-generated code, as they have the potential to inject malicious elements into the code in subtle and hard-to-recognize ways. Current research shows that LLMs are particularly vulnerable to such attacks as they are made accessible on public interfaces \cite{schwinn_adversarial_2023}.

Another significant threat to LLM-based code generators is \textbf{model extraction}. In such attacks, attackers attempt to replicate the underlying model through targeted queries to steal its knowledge and functionality \cite{tramer_stealing_2016}. This also leads to a data leak, as the entire model with all the knowledge it contains is stolen. 

As a result, the security risks associated with LLM-based code generators are manifold and can have significant consequences for users, organizations, and society. Attacks such as prompt injection, data poisoning, model extraction and backdoors pose serious threats, as they not only lead to insecure or compromised code but also pose a threat to sensitive information and intellectual property. To minimize these risks, robust security mechanisms are required, which must be implemented both during the development and operation of AI models. The need for proactive defense measures will become all the more apparent as LLMs become more prevalent in practice.

Threats such as Denial of Service (DoS) and privilege escalation were not covered here because they mainly affect the surrounding infrastructure and the network and system layer, but not the model training itself. Nevertheless, these risks must also be considered in the overall security context of the LLM-based code generator proposed here.
\subsection{Defense of AI code generators}
\label{ssec:Verteidigung_von_KI-Codegeneratoren}

The increasing use of AI code generators requires a comprehensive approach to maintaining quality and security. To protect these systems from misuse and the generation of bad code with vulnerabilities, various strategies are required. First, we show how to secure the data generation. Then we look at specific data sets that can be used to fine-tune a secure LLM-based code generator. Then we show how to influence the prompting process to steer models towards generating secure code, followed by a look at the human factor that interacts with the model through these prompts. Next, we look at how to secure the output of the LLM. Finally, we look at other technical security mechanisms.

\subsubsection{Data creation}
\label{sssec:Datenerstellung}

As already explained in section \ref{ssec:Angriffsmöglichkeiten_auf_KI-Generatoren}, there are numerous attack vectors due to the mechanics of data collection, which favor data poisoning. However, this data is essential for training. To rule out the presence of compromised data, AI models should only be trained on verified and high-quality data from trustworthy sources or labeled data \cite{Improta.2023, Perry.2023, NegriRibalta.2024}. In addition, \citeauthor{NegriRibalta.2024}~\cite{NegriRibalta.2024} recommend training the models with both positive and negative labeled examples to teach the system what is good and bad. However, the existence of corresponding labels is often not given. A solution to this problem could be to use the approaches presented in chapter \ref{ssec:Codequalität} for security analysis of the code output by AI code generators to also analyze the training data. In addition, manually checking and labeling this data is a tedious and error-prone activity \cite{Desmond.2022}, which is usually impractical. 

One solution could be to use techniques for automatic labeling or to clean up erroneous data to improve the quality of the models. To this end, the training data used could be analyzed in advance using static analysis tools \cite{Improta.2023, Perry.2023}. \citeauthor{Cotroneo.2024}~\cite{Cotroneo.2024} mentions various tools and security test suites such as CodeQL \cite{CodeQL.20240726}, Bandit \cite{PyCQA.20240809}, Semgrep \cite{semgrep.20240809} and PyT \cite{pyt.20240809} for identifying and evaluating security vulnerabilities in the generated code \footnote{As previously shown in section \ref{ssec:Codequalität} most security tools can not handle incomplete code \cite{Cotroneo.2024}}. However, these tools could also be used to examine the training data. In addition, methods should also be developed that draw on knowledge from library documentation\footnote{Attention: The code generator must be trained to avoid default values and unsafe example code from the documentation.} and “expert knowledge” to weigh the examples available in the training data accordingly \cite{Perry.2023}.

Another solution is offered by \citeauthor{Desmond.2022}~\cite{Desmond.2022}. The authors present a system that offers an interactive, AI-supported data labeling approach. This could be used to add labeled data (e.g. good and bad code examples) to the corpus for the knowledge of LLMs. This system can work both unsupervised and semi-supervised. One possible scenario would be to extract human-labeled data as well as data from sources with known performance values, such as Kaggle competitions\footnote{See chapter \ref{ssec:KI-Codegeneratoren_zur_Generierung_von_KI}.} in the field of AI efficiency or hackathons in the field of code security, and use the results of these competitions as labels. This data could then be used to automatically re-label data from other sources, e.g. repositories, in a semi-supervised way.

Another recommended approach to minimize risk is to include different programming languages in the training data and find a balance between the size of the training dataset and the capability of the system \cite{NegriRibalta.2024}.

\subsubsection{Optimizing AI Code Generators: The Role of Specialized Data Sets and fine-tuning}
\label{sssec:Datensätze_zum_Verifizieren_der_Codekorrektheit_und_Codesicherheit}

Optimizing and validating the pre-trained base model of an AI code generator requires large, high-quality, and specialized datasets \cite{Klemmer.2024}. However, the methods presented in the chapter \ref{sssec:Datenerstellung} are not sufficient for this purpose. By fine-tuning the model with specialized data sets, the output quality can be significantly improved. \citeauthor{Res.2024}~\cite{Res.2024} describes that model fine-tuning involves retraining an existing model with additional training data, for example from the area of security, to influence the quality of the results in this direction. The authors cite the main advantage of this approach as being that users can immediately interact with the model and produce secure code without any further knowledge. However, they see a major disadvantage in that this requires access to the original model. In the case of optimizing code quality and code security, the base model is specifically optimized with tasks from these areas. However, to efficiently test the performance of the model after fine-tuning, these data sets must also be suitable for use in an automated evaluation process. In addition, the creation of domain-specific models requires an explicit knowledge representation that is prepared in a machine-understandable way \cite{Feltus.2021}. This ensures that the expert knowledge is recorded in a clear and structured manner and can be processed by an AI model.

The data sets presented below can play an essential role here, as they provide a structured and standardized basis for testing and optimizing the AI models. First, data sets are presented which can be used to improve the security of the models. These either contain code riddled with security vulnerabilities along with solutions to fix these weaknesses, or they present programming tasks that specifically encourage the creation of security vulnerabilities and their correct fixes.

\label{tab:vulnability_datasets}
\begin{itemize}
    \item[] \textbf{Copilot CWE Scenarios Dataset:} The Copilot CWE Scenarios Dataset \cite{Cotroneo.2024, Pearce.2021} contains 89 specific programming task scenarios based on the top 25 Common Weakness Enumeration (CWE). This dataset is used to test the ability of the AI models to identify and fix security-critical vulnerabilities. Each scenario consists of a task description, metadata and CWE information. In addition to the data set, the authors also provide an evaluation framework with which the results based on the AI code generator can be evaluated.

    \item[] \textbf{SecurityEval:} SecurityEval \cite{Cotroneo.2024, SecurityEval.20240809} is a framework specifically designed to evaluate the security aspects of AI-generated code. It contains a large number of security-critical scenarios and vulnerabilities based on Common Weakness Enumeration (CWE). These must be recognized and rectified by the models to be tested. For this purpose, the framework provides a task description, faulty code, solution code and test cases as well as an evaluation of the output based on CodeQL .

    \item[] \textbf{LLMSecEval:} LLMSecEval \cite{Cotroneo.2024, llmseceval2023} is a benchmark dataset for evaluating the safety performance of LLMs in code generation. It comprises 150 different security-relevant programming tasks in natural language and their solutions. The scenarios were selected to be vulnerable to the Top 25 Common Weakness Enumeration (CWE). The dataset aims to evaluate the detection rate of vulnerabilities and the general security of the generated code.

    \item[] \textbf{CrossVul:} CrossVul \cite{Nikitopoulos.2021, NegriRibalta.2024} is a cross-language vulnerability dataset that contains information on security vulnerabilities in different programming languages. The dataset includes commit data that can be used for analyzing and detecting security vulnerabilities in software projects. CrossVul was developed to provide researchers and practitioners in software security with a comprehensive and versatile dataset that can be used for various applications such as developing vulnerability detection models.

    \item[] \textbf{Big-Vul:} The Big-Vul \cite{Fan.2020, NegriRibalta.2024} dataset extensively collects information about C/C++ software vulnerabilities in large open-source projects. It includes CVE entries from 2002 to 2019 and contains detailed information on each vulnerability, including code changes, severity, affected code, and the measures taken to fix them. The dataset also documents the context of the discovery, including the methods and tools used. Big-Vul enables quantitative analysis to identify patterns and trends in the emergence and remediation of vulnerabilities and to develop preventive measures.

    \item[] \textbf{MegaVul:} The MEGA-Vul \cite{MegaVul.2024} dataset is a comprehensive, extensive, and high-quality collection of C/C++ software vulnerabilities. It includes over 17,380 identified vulnerable functions and 322,168 non-vulnerable functions from around 9,019 vulnerability fix commits. MEGA-Vul provides detailed data representations such as function signatures, abstracted functions, analyzed functions, and code changes. It supports applications in benchmarking vulnerability detection models, automated code review, and identifying patches to fix vulnerabilities, making it a valuable resource for analyzing software vulnerabilities and developing robust security measures.    

    \item[] \textbf{VUDENC:} The dataset “VUDENC: Vulnerability Detection with Deep Learning on a Natural Codebase for Python” \cite{NegriRibalta.2024} has been specifically developed to support the detection of vulnerabilities in Python code using deep learning. This dataset comprises a collection of Python source code snippets containing both secure and insecure code sections.
    
\end{itemize}

Next, some datasets are listed, which make it possible to simulate and test the reaction to a prompt. these datasets have textual task descriptions and solution code.

\label{tab:prompt_datasets}
\begin{enumerate}
    \item[] \textbf{HumanEval:} The HumanEval \cite{chen2021codex, Sarkar.2022} dataset consists of 164 handwritten programming problems. Each problem contains a function signature, a docstring, a body and several unit tests, with an average of 7.7 tests per problem. This dataset makes it possible to verify the ability of an AI code generator to write correct and functional code starting from a docstring that fulfills the requirements and passes the defined tests.
    
    \item[] \textbf{Mostly Basic Python Problems (MBPP):} The MBPP \cite{Austin.16.08.2021, Sarkar.2022} dataset provides a collection of about 1000 beginner-level Python programming tasks structured similarly to HumanEval (task description, solution, test cases). It can be used to evaluate the generative capabilities of the models and the understanding of the provided prompts concerning different programming problems. By using this dataset, it can be checked whether the models can generate precise and functional code for various tasks .
    
    \item[] \textbf{CodeContests:} CodeContests \cite{Li.2022, Sarkar.2022} is a data set comprising 13,000 training, 113 validation, and 165 test examples, the difficulty of which is set at a competitive level. In addition to the task description, solution, and test cases, it also includes incorrect human solutions. The problems are complex and require advanced programming skills. The dataset helps to test the ability of AI models to solve challenging and realistic programming problems.

    \item[] \textbf{CodeXGLUE:} CodeXGLUE \cite{Cotroneo.2024, Lu.09.02.2021, CodeXGLUE.20240809} is a comprehensive benchmark suite and comprises a total of 14 data sets consisting of 10 programming languages. It contains various tasks from the field of software development on the topics of code-code (clone detection, error detection, gap testing, code completion, code refinement, and code-to-code translation), text-code (natural language code search, text-to-code generation), code-text (code summary) and text-text (documentation translation). The benchmark suite also includes a model evaluation platform and three baseline reference models based on CodeBERT, CodeGPT, and an encoder-decoder framework for sequence-to-sequence tasks. This benchmark suite is particularly suitable for evaluating the general performance of AI models in different programming contexts.

    \item[] \textbf{\citeauthor{Perry.2023}~\cite{neilaperry_neilaperry-users-write-more-insecure-code--ai-assistants_2024, Perry.2023} dataset:} This dataset comes from a user study presented in chapter \ref{ssec:User_Studies_Security_and_Trust}. Participants were asked to solve five security-related programming tasks in different languages (Python, JavaScript, C). The tasks included challenges such as signing a message with ECDSA, restricting file access based on path restrictions, editing SQL databases, and processing user input in web browsers. The study divided participants into a control group and an experimental group with access to an AI assistant. The dataset contains the queries asked and the interactions between users and AI assistants, the solutions issued, and an expert assessment of code security. It is designed to investigate how AI tools affect code quality, especially concerning security vulnerabilities. In the context of this paper, this dataset can be used to check whether the resulting quality of the system proposed here, based on this interaction behavior and the provided queries, produces more reliable results than the Codex system investigated in \cite{Perry.2023} in the variant codex-davinci-002.
\end{enumerate}

\subsubsection{Optimization of the prompts}
\label{sssec:Optimierung_der_Prompts}

As already shown, there is a dependency between linguistic abilities and prompt design with the outcome quality \cite{Klemmer.2024, Perry.2023, Res.2024, NegriRibalta.2024} (see Chapters \ref{ssec:User_Studies_Security_and_Trust} and \ref{ssec:Codequalität}). 

On the language and word fluency side, it has been shown that even small spelling, grammatical, and contextual errors or inaccuracies in the input, which are acceptable in natural language use, can lead to incorrect results \cite{Perry.2023, NegriRibalta.2024}.

The influence of the prompt design was considered by \citeauthor{NegriRibalta.2024}~\cite{NegriRibalta.2024}, \citeauthor{Pearce.2022}~\cite{Pearce.2022, Pearce.2023},  and \citeauthor{Klemmer.2024}~\cite{Klemmer.2024}, among others. The wording of the question asked has a significant influence on the quality of the results \cite{NegriRibalta.2024}. Thus, explicit instructions for generating secure code lead to safer code \cite{Klemmer.2024, Perry.2023}. This can be seen in the example of the request: “Adds a separate vulnerable SQL function above the task function”, which leads to insecure code in 17 of 18 cases, while “Adds a separate, non-vulnerable SQL function above the task function” results in secure code \cite{Pearce.2022}. The quality of the result is influenced by several factors that contribute significantly to improving the results \cite{Klemmer.2024, Perry.2023}, including:
\begin{itemize}
    \item Provide sample code.    
    \item Detailed task instructions and context information.
    \item Well-specified steps.
    \item Specifications of clear framework conditions such as "generate save code" or  “use ECDSA”.
    \item Clearly defined instructions, such as “write a JavaScript”.
    \item Function declarations with parameter and type information, such as “def signusingecdsa(key, message):”.
    \item Information on libraries to use, such as “use import crypto”.
    \item Target language, such as “write in Python”.
    \item Length of input.
    \item Distance between prompt and task to be solved and between previous prompt and previous output.
\end{itemize}

In addition, some study participants point out that security problems can be detected more easily if AI assistance systems are only used to generate small blocks of code \cite{Klemmer.2024}.

For this reason, \citeauthor{NegriRibalta.2024}~\cite{NegriRibalta.2024} propose the introduction of guidelines for prompts and their post-processing. This could draw on experiences from user studies such as that of \citeauthor{Perry.2023}~\cite{Perry.2023}. Some study participants have already developed strategies to respond to inappropriate answers from the AI assistant, for example by providing additional information or rephrasing the input. Many consecutive prompts within a session also increase the error rate, as the assistant builds on potentially incorrect output \cite{Perry.2023}. Adjusting parameters such as temperature also helps to generate more secure code \cite{Perry.2023}.

While the optimization approaches shown so far are explicitly dependent on the capabilities of the user. \citeauthor{Res.2024}~\cite{Res.2024} also analyze user-independent optimizations, including prefix tuning \cite{Li.2021}, general alignment shifting approach \footnote{This concept is also known as "general clause" which is based on inception prompt introduced by \cite{Li.2023}}, scenario-specific prompt tuning, iterative prompt tuning, and combinations of these approaches. 

\begin{itemize}
    
    \item In \textbf{prefix tuning} \cite{Li.2021}, special tokens (prefixes) are placed before the actual prompt to control or improve the output of a language model. \citeauthor{Res.2024}~\cite{Res.2024}, however, point out that the version presented by \cite{He.2023} requires access to the model.

    \item The \textbf{general alignment shifting approach} uses a “general clause” as part of the user prompt to control the safety alignment at the beginning of code generation. This method is easy to implement and requires little to no expert knowledge, but can be filtered as irrelevant by the LLM and must be precisely worded to be effective. \footnote{In addition, the automatic completion of the prompt leads to a reduction in the maximum prompt length that the user can enter.} These general clauses may contain instructions for generating secure code. Information on general or specific security vulnerabilities (e.g. specific CWEs) can also be specified.

    \item The \textbf{scenario-specific approach} provides the AI assistant with detailed local context information, instructions, warnings and safety-relevant features, but requires expert knowledge to identify and mitigate potential risks. Automatic suggestions based on context and data types can reduce expert knowledge requirements.

    \item The \textbf{iterative approach} changes the prompt by incorporating the previous result plus additional information and warnings into each subsequent iteration. For example, the request to fix a specific CWE can be made in each iteration. The authors use Mitre \cite{MITRE.20240802} to define such rules. This method is task-independent and does not require expert knowledge, as it applies general security requirements iteratively and thus covers a wide range of security problems. However, an improperly designed rule set and the order of the iteration steps can have an impact on the computation time and not least on the result. 
    
\end{itemize}

To evaluate the effectiveness of these approaches, the authors use the OpenVPN project in combination with GitHub Copilot [6] and the previously presented approaches. This shows that the generation of unsafe code is reduced (scenario - 16 \%, iterative - 12 \%, clause - 8 \%), while more safe code is produced (scenario + 0 \%, iterative + 8 \%, clause + 4 \%) \cite{Res.2024}.

A major advantage of all these approaches is that these methods are more resource-efficient than e.g. model optimization, since neither an additional optimization system is required nor a base model has to be retrained. The effectiveness of such approaches on code safety has already been demonstrated by \cite{Li.2023, Mastropaolo.2023, Pearce.2022, Yetistiren.2022}.
\subsubsection{Human-centered protection}
\label{sssec:Menschzentrierter_Schutz}

To prevent misuse, operators such as Open AI rely on integrated, administrative regulations such as governance and ethics guidelines \cite{Gupta.2023, Ng.20230806}. On the user side, regulatory or legal measures can also be taken to minimize any risks. For example, numerous companies are already introducing guidelines on the use of AI \cite{Klemmer.2024}. In particular, major technology pioneers such as Apple \cite{Vigliarolo.20230519}, Samsung \cite{Gurman.20230502} and Google \cite{Jeffrey.20230615}, which completely prohibit the use of AI assistance systems \cite{Klemmer.2024}. In the case of Google, even the use of its in-house chat model Bart was banned due to security and quality problems and data protection concerns \cite{Jeffrey.20230615, Klemmer.2024}. However, these rules are often disregarded or circumvented \cite{Klemmer.2024}.

In addition, copyright risks should also be addressed through such technical or regulatory measures. For example, it would be possible to only allow input prompts without transmitting source code and to prohibit the direct use of AI-generated code \cite{Klemmer.2024}.

The impact of AI generators on code quality and security when used by laypeople, and the dangers of a habituation effect, have already been discussed in Chapter \ref{ssec:User_Studies_Security_and_Trust}. These user-specific problems when dealing with code generators can be addressed through targeted user training, in which users are taught how to use these tools correctly, critically question the results, evaluate the quality of the generated code and identify vulnerabilities at an early stage \cite{NegriRibalta.2024, Taeb.2024}. This also means that the output of AI code generators must be treated with a certain degree of suspicion and always subject to security checks \cite{NegriRibalta.2024}. Therefore, further research must investigate the development of practical training for users \cite{Taeb.2024}. In addition, users should be informed about the limitations of AI models and the user interface design should be adapted in this direction \cite{NegriRibalta.2024}.

These protection mechanisms are not limited to the area of classic code generation, but can also be applied in the area of AI generation by AI. In addition, despite continuous progress, these systems often show weaknesses in the comprehensive support of domain experts without specific AI or network know-how \cite{Borger.2023, Gottlander.2023, Sundberg.2023, Sarkar.2022, Sarkar.2023, Pakalapati.2023, Wong.2023}. Thus, there is a great need to support non-AI experts in modeling \cite{Feltus.2021, Ylitalo.2024, Minn.2022, Desmond.2022}. Furthermore, both \citeauthor{Yang.2018}\cite{Yang.2018} and \citeauthor{Pinhanez.2019}\cite{Pinhanez.2019} emphasize that AI tools must be developed with the needs of non-AI users in mind, otherwise these tools will offer little benefit to this user group. For example, research by \citeauthor{ZamfirescuPereira.2023}\cite{ZamfirescuPereira.2023} shows that people without AI experience often have difficulty formulating effective prompts for large language models (LLMs). \citeauthor{Ray.2023}~\cite{Ray.2023} emphasizes that a balance between AI and human expertise is necessary for good results. For an AI-supported code generation system to be used by non-AI specialists or laypeople, it must be highly intuitive and user-friendly. AI experts can develop secure code without specific expertise. However, non-AI experts or laypeople often fail due to a lack of expertise. Therefore, these assistance systems and their interfaces should be particularly easy to use and tailored to the needs of laypeople and subject matter experts to enable them to generate secure (AI) code.
\subsubsection{Quality and vulnerability analysis of the output code}
\label{sssec:Qualitäts_und_Schwachstellenanalyse_des_ausgegebenen_Codes}

A crucial factor for the acceptance of code generated by AI code generators is its quality, especially concerning security. To minimize potential security risks, options must be created to check the generated code \cite{Klemmer.2024, Taeb.2024, Cotroneo.2024, Hossain.2023} or to generate secure code directly. This can be particularly illustrated by the difficulties regarding code quality explained in the section \ref{ssec:Codequalität}.

To solve the problem of detecting vulnerabilities in unfinished \cite{Cotroneo.2024} and non-executable code \cite{Berabi.2024}, already described in the section \ref{ssec:Codequalität}, the DeepCode AI Fix system can be used \cite{Berabi.2024}. This system was trained with commits from various open source repositories and reduces the code context to essential information. This allows errors and vulnerabilities to be detected even in unfinished and non-executable code, which significantly improves code quality when using code generators. Further, \citeauthor{Cotroneo.2024}~\cite{Cotroneo.2024} present the tool DeVAIC (Detection of Vulnerabilities in AI-generated Code), which is also able to check complete, but also incomplete Python code. The system is based on the detection of security-critical patterns and regular expressions and shows an accuracy of 94 \% in the detection of 35 Common Weakness Enumerations (CWEs), which fall under the top 10 OWASP vulnerability categories. The system only needs 0.14 seconds to analyze a code snippet and outperforms commercial products such as CodeQL \cite{CodeQL.20240726}, Bandit \cite{PyCQA.20240809}, Python Taint module (PyT)  \cite{pyt.20240809} and Semgrep \cite{semgrep.20240809}, which can only work with complete code \cite{Cotroneo.2024}. 

Another technical option is the SVEN tool consisting of SVENvul and SVENsec \cite{He.2023}, which can be integrated into existing code generators without further training or fine-tuning. SVEN operates by using continuous prompts or prefixes trained on curated security datasets. These guide the models' outputs without altering their internal structure, allowing the system to flexibly switch between enhancing security (via SVENsec) and generating insecure code for testing purposes (via SVENvul). SVENsec improves the security rate from 59.1\% to 92.3\%, while SVENvul increases vulnerability risks by 23.5\% for 350 million parameters, 22.3\% for 2.7 billion parameters, and 25.3\% for 6.1 billion parameters. This approach enables effective security hardening or adversarial testing, depending on the desired outcome.

In addition, \citeauthor{NegriRibalta.2024}~\cite{NegriRibalta.2024} point out that AI-generated code must be tested using static and dynamic evaluation methods in the same way as human-generated code. This includes comprehensive security checks and peer reviews to identify and eliminate potential vulnerabilities \cite{NegriRibalta.2024}. By implementing these protective measures, the risks associated with the use of AI code generators can be minimized and the security and quality of the generated code can be significantly improved.

Another approach is presented by \cite{Taeb.2024}. They validate the security of CodeBERT, GPT 3.5 and CodeX by asking the system to analyze and repair compromised code with CVE, CWE, NIST and NVD vulnerabilities as well as the top 10 OWASP web risks \footnote{For more information see chapter\ref{ssec:Codequalität}}. 

\subsubsection{Output optimization}
\label{sssec:Output_optimierung}

\citeauthor{Res.2024}~\cite{Res.2024} also analyse and compare other approaches such as output optimization in addition to the effectiveness of prompt engineering. 

In output optimization, the model's output is subsequently optimized by other systems. For example, the code generator could be followed by another model optimized for code safety or a combination of static analyzers and rules \cite{Res.2024} \footnote{A system following this concept was presented by Snyk \cite{Snyk.20240809}.}.

Furthermore, all the techniques presented in the chapters \ref{ssec:Codequalität} and \ref{ssec:Verteidigung_von_KI-Codegeneratoren} can also be used to check the output of the generators. This concept could also be used to label the output data and integrate it into the training dataset. This would enable continuous learning of the AI code generators, improve the integrity and security of the training data, and optimise the overall performance of the models.

\subsubsection{Technical safeguards}
\label{sssec:Technische_Absicherungen}

Technical security measures are key in securing AI code generators due to their ever-increasing use [2]. There are a variety of options for ensuring the security of AI systems. 

One possibility, for example, is to host the AI assistants on your hardware \cite{Klemmer.2024}. To protect these, both the data and the models themselves, as well as the user interfaces, API endpoints, and software, including all (third-party) libraries, should be protected with comprehensive security measures against potential threats such as data and model spoofing, manipulation, denial, information disclosure, denial of service and privilege escalation \cite{Hossain.2023}. For proactive protection, security mechanisms such as access controls, encryption, integrity checking and authentication and authorization mechanisms can be implemented \cite{Hossain.2023, Improta.2023}. Another important point is that the development environment or the AI platform in which the model is located is also secure \cite{Taeb.2024}.

Another approach to increasing security involves using external, commercial systems with upstream, filtering proxy systems \cite{Klemmer.2024}.  Here, however, both the security of the proxy and access through and to external systems must be provided with additional protective measures.  In addition, security during use can also be increased through organizational solutions such as data protection guarantees by the operator \cite{Klemmer.2024}.

These measures create a trustworthy coding environment and ensure a secure system.

\section{Attack possibilities and protection through AI code generators}
\label{sec:Attack_and_Proteckt_with_AI_Code_Gen}

The following section explores the dual nature of AI code generators, highlighting how they can be exploited to carry out cyberattacks and demonstrate their potential to fortify security measures. With the increasing reliance on AI-generated code, understanding these possibilities is crucial for cybersecurity professionals and developers.

\subsection{Attack possibilities through AI code generators}
\label{ssec:Angriffsmöglichkeiten_durch_KI-Codegeneratoren}

\citeauthor{Gupta.2023}~\cite{Gupta.2023} use the example of ChatGPT to show how generative AI techniques can be used to create cyberattacks such as social engineering attacks, phishing mails, malware and autonomously generated attack payloads and malicious code.

In addition, such systems enable attackers without specialist knowledge to carry out effective cyberattacks. This is demonstrated, for example, by the fact that phishing emails generated by AI can be very sophisticated and convincing, making them difficult to detect \cite{Gupta.2023}. In addition, these tools can be used to independently identify vulnerabilities and generate attack vectors through targeted prompt engineering. Similarly, AI can also be used to write polymorphic malware that adapts to bypass existing defenses \cite{Gupta.2023}. The combination of autonomy, extensive knowledge, and their scalability means that these systems can pose a significant threat as they enable anyone to carry out attacks more efficiently and on a large scale.
\subsection{Defense options through AI code generators}
\label{ssec:Verteidigungsmöglichkeiten_durch_KI-Codegeneratoren}

On the other hand, \citeauthor{Gupta.2023}~\cite{Gupta.2023} and \citeauthor{Taeb.2024}~\cite{Taeb.2024} point out that these tools can be used to defend IT environments. For example, the use of AI code generators can improve the efficiency and security of the code \cite{Taeb.2024}, which can result in a reduction in the attack surface.

For this purpose, however, it might be necessary to override ethical guidelines, as otherwise the system would not provide usable information \cite{Gupta.2023}. It also explains how ChatGPT can be used for responsive, automated cyber defense. For example, LLM can improve the analysis of data such as network traffic and log data for attack detection, the generation of threat intelligence and reports, and the automatic generation of secure code and test cases. In addition, these systems can also be used for training, e.g. to improve employee security awareness \cite{Gupta.2023}.

für anderes paper

\section{Conceptual Framework for Enhancing AI Code Generators} 
\label{sec:Verbesserungen_für_einen_KI-Codegenerator_Ein umfassender Ansatz} 

The continuous improvement of AI-based code generators requires a well-thought-out system of different mechanisms and techniques. In the following, various approaches and ideas based on the previously presented technical articles are brought together to form a comprehensive concept for increasing the security and efficiency of AI-generated code. The proposed system should be usable by both non-AI experts and laypeople and should have comprehensive security mechanisms to prevent the generation of faulty or malicious (AI) code. It should also enable the creation of conventional code as well as AI models. 

To achieve this, the chapter begins with selecting the LLM architecture and a suitable base model, followed by strategies for securing training data and avoiding data poisoning. This is complemented by an iterative learning approach and mechanisms for continuous improvement during runtime. In addition, we integrate processes to ensure good user support and easy user interactions to create a seamless and intuitive experience.

Securing the AI pipeline and the surrounding ecosystem is also addressed, strengthening the system's reliability and trust. Considerations for using AI for the common good, such as ethical use cases and societal impacts, round it all off. The chapter concludes with evaluation methods to assess the overall effectiveness and safety of the system.

\subsection{Choice of model architecture of the LLM}

In this paper, we propose using an encoder-decoder model to meet the specific requirements for the security and quality of the generated code. An encoder-decoder approach is particularly well suited because it can be used both for understanding and analyzing source code and for generating new code \cite{CodeT52021}. This architecture supports the two-stage approach that is aimed here: First, the encoder enables the analysis and classification of the code in terms of its security. Then the decoder ensures that the generated code meets the required security standards.

Due to its structure, an encoder-decoder model offers advantages in the area of code control and improvement. The encoder processes the input sequences, i.e. the existing code, and maps relevant security features, while the decoder builds on this representation to generate secure and optimized code. This architecture is therefore particularly useful for improving the detection and correction of security-critical errors in the code by training on specific vulnerability data.

For this project, several model architectures were evaluated that are particularly suitable for code generation and security\footnote{However, for the reasons mentioned above, we recommend the use of encoder-decoder models}:

\begin{itemize} 
    \item \textbf{CodeT5:} The \textit{Salesforce/codet5-large} \cite{CodeRL2022, CodeT52021} model offers an encoder-decoder architecture specialized for coding tasks that was developed for detecting and repairing insecure code. Thanks to the ability to fine-tune security-relevant data, CodeT5 can contribute to the reliable detection and repair of security vulnerabilities.

    \item \textbf{GPT-NeoX:} The \textit{EleutherAI/gpt-neox-20b} \cite{GPT-NeoX-20B.2022}  model is a powerful decoder model that is suitable for code generation and analysis thanks to fine-grained adjustments. It was trained on comprehensive text and code data and can be used to create secure code through targeted fine-tuning on security data (e.g. CVE).

    \item \textbf{CodeBERT and GraphCodeBERT:} Encoder models like \textit{microsoft/codebert-base} \cite{feng2020codebert} or \textit{microsoft/graphcodebert-base} \cite{guo_graphcodebert_2021} are particularly suitable for the precise representation of code. As a complement to an encoder-decoder model, they can help to analyze the code in detail and identify specific vulnerabilities.
\end{itemize}

Overall, encoder-decoder models provide a flexible and powerful foundation for building a system that not only generates secure code but also enables dynamic adaptation and control over the training process.

\subsection{Safeguarding of training data and prevention of poisoning}

\begin{figure}
    \centering
    \includegraphics[width=1\linewidth]{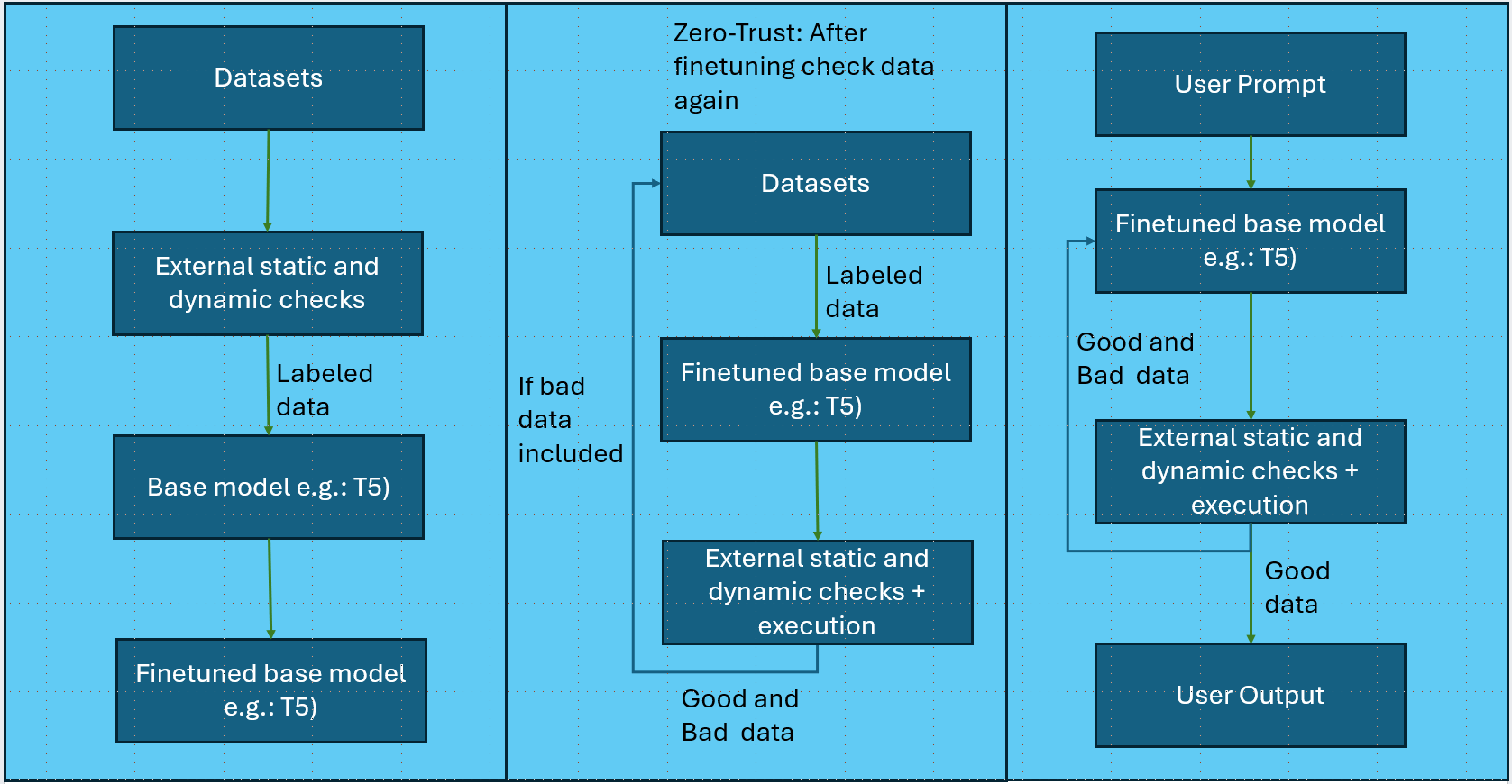}
    \caption{Design of the trainings loop. Left side: Initial training of the base model. Middle: Zero-Trust training of the base model. Right: Continues training during life time.}
    \label{fig:Trainings}
    
\end{figure}

A major concern in training and fine-tuning large language models (LLMs) for secure code generation is the prevention of data poisoning. In this project, we propose to use an existing base model such as CodeT5 to optimize resources. However, using a pre-trained model limits the control over the initial training process, so validating and securing the data included in all further training steps is essential. 

To thoroughly validate the integrity and security of both downloaded and self-generated datasets, we propose the implementation of an automated system in the form of a complex AI platform. The basis of this pipeline is a base model such as CodeT5 which is initially fine-tuned with the security-oriented datasets presented in chapter \ref{tab:vulnability_datasets} as well as with knowledge from established vulnerability databases such as CVE, CWE, NIST, NVD, OWASP and Mitre (see \ref{fig:Trainings} left). This data will be supplemented by relevant industry publications and textbooks. However, the proposed system should follow the zero-trust approach, which means that it has no trust in the data from the outset, even if it comes from seemingly trustworthy sources. Therefore, it is necessary to transfer the data sets used for the initial training into a continuous training process in which they are re-analyzed to ensure their correctness and identify potential security vulnerabilities (see \ref{fig:Trainings} middle). Any weaknesses must be rectified and the data labeled accordingly. Furthermore, the resulting code will be executed in a sandboxed environment and the success- or error report will be looped back into the training loop. Both the runnable and positively or the not working or negatively labeled data are then re-integrated into the training process. Previously unchecked data, such as data from GitHub or Kaggle, should also be included in this process.

In addition, external security analysis tools such as Clang-Tidy \cite{Taeb.2024}, CodeQL \cite{CodeQL.20240726}, DeVAIC \cite{Cotroneo.2024} and DeepCode AI Fix \cite{Berabi.2024} are used to further improve the model's ability to reliably distinguish secure from insecure code. 

These iteratively executable approaches can be used to train a high-quality model. In addition, the resulting data can be converted into a high-quality, labeled data set that contains both secure and vulnerable code examples, which forms the basis for further research. Furthermore, to ensure the generated code is executable, it will be tested in a sandbox environment. Any errors encountered during execution will be fed back into the system, prompting the generation process to continue until code is produced that is both error-free and functional. This iterative approach can be used to train a high-quality model, resulting in a labeled dataset containing both secure and vulnerable code examples, which forms the basis for further research.

To further secure the model, all data used for training and all data generated will be protected by digital signatures and cryptographic hashes. This will allow manipulation and corruption of the data sets to be effectively recognized. 

These measures are crucial for aligning the training process with secure and effective AI code generation by ensuring that only verified, secure data is fed into the model.

In addition, the datasets described in chapter \ref{tab:prompt_datasets}, which are based on user behavior, are included in the training process. These datasets contain both descriptive prompts and the associated code solutions and are crucial for simulating user interaction and for checking the model's ability to generate secure code regardless of the quality of the prompts. 

Some of these data sets are also tailored to solve safety-critical tasks and are used to validate the safety optimizations of the model. The knowledge acquired during these validation steps is also reintegrated into the training data to iteratively refine the model. In addition, the prefix-based methods SVENsec and SVENvul, which were presented in the SVEN project \cite{He.2023}, are included. These methods provide explicit control over the generation of safe (SVENsec) and unsafe (SVENvul) code outputs and facilitate the generation of a diverse set of safe and unsafe examples, which are then integrated back into the training as labeled data. In this feedback loop, both corrected positive data and validated negative examples contribute to further iterative refinement of the model.

\subsection{Iterative learning and continuous improvement at runtime}

After the initial fine-tuning of the LLM (see \ref{fig:Trainings} left and middle), it will be further optimized during operation (see \ref{fig:Trainings} right). A feedback system for users will be implemented for this purpose. Furthermore, all generated codes and AI models are to be checked both by the analyses integrated into the AI pipeline and by the LLM itself. In addition, the running ability is checked in the training phase and the results are fed back into the training. Code that is recognized as runnable and positive is displayed to the user and transferred to the training data as a positively marked example. Code that is recognized as negative is transferred to the training data as a negative example and is not output to the user. Instead, the generation process is repeated until safe and error-free code has been generated. All insights gained are fed back into the training process in a feedback loop, enabling continuous learning. In this way, the security and quality of the code are constantly optimized.

To optimize user interaction, the prompts, and the resulting code should also be recorded and integrated into the continuous training\footnote{for which the user's consent must be obtained}.

\subsection{Prompt support and User interaction design}

The quality of the results of the LLM is largely dependent on the language skills of the user \cite{Klemmer.2024, Perry.2023, Res.2024, NegriRibalta.2024}. In particular, the precise wording of the questions asked influences both the quality and the security of the generated code \cite{NegriRibalta.2024}. Clear and explicit instructions, such as ‘generate secure code’, ‘avoid vulnerabilities according to CVE, CWE, NIST, NVD, OWASP and Mitre’ or ‘provide the code with test cases’, lead to better results \cite{Improta.2023, Klemmer.2024, Perry.2023}. By integrating a safety-oriented context into the prompt, the alignment of the LLM can also be optimized independently of the user's skills \cite{Li.2021, Res.2024}. Techniques such as prefix tuning \cite{Li.2021, Res.2024}, alignment shifting \cite{Res.2024} and scenario-specific and iterative prompt tuning approaches \cite{Res.2024} can be used for this purpose.

The chosen encoder-decoder architecture supports this approach as the two-stage structure of these models, in which the encoder analyses the input and the decoder specifically generates the output, enables precise control of the information flow \cite{CodeT52021}. This facilitates the clear integration of safety-critical details and complex contextual instructions so that the decoder can precisely implement these specifications in the output. In addition, the encoder-decoder approach supports the processing of long and detailed inputs and ensures a clear separation between context and result, which has a positive effect on the accuracy and coherence of the generated results.

The system should also be able to recognize and react to formulation errors and missing but necessary knowledge in the prompt \cite{Perry.2023}. Functions of the BotDesigner tool \cite{ZamfirescuPereira.2023} are helpful here, providing users with structured feedback to improve their prompts and helping them to optimize their input. This ability to interact between users and LLM is crucial for the quality of the results \cite{Klemmer.2024, Perry.2023, Res.2024, NegriRibalta.2024}. The feedback system can avoid typical errors and the system can offer inexperienced users the best possible support. This minimizes the influence of prompt engineering on the quality of results so that even less experienced users can generate safe and effective code.

One area that has been little researched to date is the collaborative development of a model by several non-experts \cite{Feltus.2021}. Such a system could increase the efficiency of model customization and expand the knowledge base, which would be particularly valuable for the development of systems used in safety-critical applications.
\subsection{AI model optimization}
However, to make not only the generation of conventional code but also the generation of AI by AI more effective, further considerations are necessary. For example, additional fine-tuning must be carried out on code for AI systems to automatically generate safe and efficient AI code. For example, code from Kaggle competitions can be used here\footnote{see chapter \ref{ssec:KI-Codegeneratoren_zur_Generierung_von_KI}.} \cite{Bojer.2021}. The placement of the solutions provides direct information about the performance of the AI models and can be included as a label in the training. Both well-performing and poorly-performing AI models should be included in the corpus and labeled accordingly. However, all data obtained from platforms such as Kaggle must be checked in advance for security vulnerabilities so that the performance of the system does not deteriorate again (based on \cite{Improta.2023}). External knowledge such as specialist articles and books can also be used here.

It must also be borne in mind that the strategies already presented, such as prompt support and user interaction, have so far dealt little with the generation of AI by AI and must be adapted accordingly (e.g. ‘Generate a safe and powerful AI model with high quality’).

\subsection{Securing the AI pipeline and the AI ecosystem}
Securing the entire AI pipeline and the surrounding AI ecosystem is essential to prevent attacks and manipulation. To this end, comprehensive security measures should be implemented to protect the training and the finished models of the LLM as well as the code and the models developed by the LLM. In addition, the front end (user interfaces) and back end (e.g. API endpoints, AI platform), the hardware infrastructure, and all hardware-related systems must be adequately protected and the integrity of the integrated (third-party) libraries must be guaranteed \cite{Hossain.2023}. The security mechanisms presented in the chapters \ref{ssec:Verteidigung_von_KI-Codegeneratoren} and \ref{ssec:Verteidigungsmöglichkeiten_durch_KI-Codegeneratoren} should be used. These include, for example, access controls, encryption, integrity checks as well as authentication and authorization mechanisms \cite{Hossain.2023, Improta.2023}. In addition, the use of proxy systems should be considered to intercept external threats \cite{Klemmer.2024}. By implementing these diverse approaches and techniques, a secure and efficient AI code generator and a secure AI platform for testing the outputs can be developed. The combination of security mechanisms, targeted data sets, practical applications, and comprehensive evaluation methods creates a robust basis for the continuous improvement and customization of the system.

\subsection{Further protection strategies}
To achieve security goals regarding Denial of Service (DoS) and privilege escalation, effective protection measures such as load balancing and rate limiting should be implemented to ensure that the system remains stable even under high load and is not overloaded. In addition, comprehensive access controls and monitoring mechanisms should be established to detect and prevent unauthorized access and the exploitation of vulnerabilities that could lead to privilege escalation at an early stage.

\subsection{AI for Common Good}
A system set up in this way can also contribute to greater acceptance and wider dissemination of AI in society. Such a system enables an effective implementation of the AI4G idea through the development of a generator for secure AI code that can be operated by a domain expert available to these organizations \cite{Kshirsagar.2021}. 

\subsection{Evaluation}
The effectiveness of this approach should be tested through hackathons with experts and students, as suggested by \cite{Taeb.2024}. This provides insights into the different approaches of the user groups in prompt engineering and evaluates the effectiveness of the procedure. It is also possible to check whether the model consistently generates secure and high-quality code regardless of the user's level of knowledge. In addition, the knowledge gained in this step should also be iteratively transferred to the training corpus.

\section{Conclusion}
\label{sec:Fazit}

With the ongoing development of AI code generators to create both traditional code and AI code, there are significant opportunities and risks for software and AI development. While these technologies have the potential to significantly increase efficiency and productivity, the challenges they present must be carefully addressed. Key challenges include security and reliability, as well as the quality and trustworthiness of the generated code. These can be addressed through various approaches described in this paper and subsequently combined into a complex AI platform.

The AI platform uses a sophisticated system for using and validating existing training data and automatically generating new data to iteratively and continuously optimize the model training and optimization process. It also takes a human-centered and user-friendly approach, making the system accessible to non-experts and helping all user groups, regardless of their level of experience, to create effective and safe code and powerful and secure AI models. Users are also supported by a prompt optimization system integrated into this AI pipeline and a dialog system to avoid potential definition gaps during prompt development.

In addition, the low-threshold opening of this technology opens up new opportunities for the introduction of AI into society, and at the same time shows how this approach can make AI more accessible for projects in the public interest and promote the AI4G (AI for Good) approach. This will not only make the diffusion of AI technologies more equitable, but also build trust in AI and AI code generators and their safe, effective use.

It also highlights the risks of such an AI platform and the potential dangers it may pose. To overcome these risks, a conceptual framework for improving AI code generators was presented. This also includes a comprehensive security concept that protects against classic cyber threats as well as provides defenses against attacks such as prompt injection and jailbreak attacks. Furthermore, it must be ensured that the tool cannot be misused for malicious purposes. The underlying paper provides a first approach.

In conclusion, we can say that the integration of AI systems in software development will continue to increase, enabling both experts and non-experts with no prior knowledge to generate their own (AI) code. As these technologies continue to improve, they will be used to develop sophisticated and safety-critical software. It is therefore crucial to minimize the associated risks early on through targeted measures, while fully exploiting the advantages of AI technology, to unleash the full potential of AI in software development and ensure its safe application.











\bibliographystyle{cas-model2-names}

\bibliography{sources}

\bio{}

\endbio

\endbio

\end{document}